\pdfoutput=1

\documentclass[sigconf,natbib,9pt]{acmart}

\copyrightyear{2017}
\acmYear{2017}
\setcopyright{acmlicensed}
\acmConference{CIKM'17 }{November 6--10, 2017}{Singapore, Singapore}\acmPrice{15.00}\acmDOI{10.1145/3132847.3132979}
\acmISBN{978-1-4503-4918-5/17/11}

\usepackage[subtle, wordspacing=normal, tracking=normal, bibnotes, charwidths=normal, leading, indent, lists, paragraphs=normal, mathspacing=normal, bibliography=normal]{savetrees}
\usepackage[ruled,ruled,noend,noline,slide]{algorithm2e}
\SetAlFnt{\small}
\usepackage{array}
\usepackage{booktabs}
\usepackage{caption}
\usepackage{dsfont}
\usepackage[inline]{enumitem}
\usepackage{etoolbox}
\usepackage{flushend}
\usepackage{framed}
\usepackage{grffile}
\usepackage[np, autolanguage]{numprint}
\usepackage{paralist}
\usepackage[nodisplayskipstretch]{setspace}
\usepackage{subcaption}
\usepackage{tabularx}
\usepackage{tikz}
\usetikzlibrary{arrows, arrows.meta, calc, decorations.markings, decorations.pathreplacing, positioning, shapes}
\usepackage{varwidth}

\tikzset{>=latex}
\SetKw{Continue}{continue}

\setcitestyle{square,comma,numbers,sort&compress,sectionbib}

\makeatletter
\renewcommand\paragraph{\@startsection{paragraph}{4}{\parindent}%
  {-.10\baselineskip \@plus -0.10\p@ \@minus -.1\p@}%
  {-3.5\p@}%
  {\bfseries\@adddotafter}}
\makeatother

\newcommand{\RetrievalSystem}{R}

\newcommand{\Length}[1]{\left|#1\right|}

\newcommand{\Time}{t}

\newcommand{\Threads}{C}
\newcommand{\Thread}{\MakeLowercase{\Threads{}}}

\newcommand{\Users}{U}
\newcommand{\User}{\MakeLowercase{\Users{}}}

\newcommand{\AttachableEntities}{E}
\newcommand{\AttachableEntity}{\MakeLowercase{\AttachableEntities{}}}

\newcommand{\UserTimeConstraintedAttachableEntities}[1][\CurrentTime{}]{{\AttachableEntities{}_{\User{}, {#1}}}}

\newcommand{\Messages}{M}
\newcommand{\Mailbox}{\Messages}
\newcommand{\Message}{\MakeLowercase{\Messages{}}}
\newcommand{\MessageTime}{{\Time{}}_{\Message{}}}

\newcommand{\RequestMessage}{\Message{}_{\text{req}}}
\newcommand{\ReplyMessage}{\Message{}_{\text{res}}}

\newcommand{\RequestReplyPair}{\langle\RequestMessage{}, \ReplyMessage{}\rangle}

\newcommand{\TargetAttachableEntity}{\AttachableEntity{}_{actual}}

\newcommand{\RequestEntityPair}{\langle\RequestMessage{}, \TargetAttachableEntity{}\rangle}

\newcommand{\RequestMessageCandidateQueries}{\CandidateQueries{}_{\RequestMessage{}}}

\newcommand{\Queries}{Q}
\newcommand{\Query}{\MakeLowercase{\Queries{}}}

\newcommand{\Documents}{D}
\newcommand{\Document}{\MakeLowercase{\Documents{}}}

\newcommand{\FrequencyFn}{\#}

\newcommand{\Apply}[2]{#1\left(#2\right)}
\newcommand{\Prob}[2][P]{\Apply{#1}{#2}}
\newcommand{\CondProb}[3][\Prob]{#1{#2 \mid #3}}

\newcommand{\CurrentTime}{\Time{}^\prime{}}

\newcommand{\CandidateQueries}{\tilde{\Queries{}}}
\newcommand{\CandidateQuery}{\tilde{\Query{}}}
\newcommand{\CandidateQueryTermBudget}{k}

\newcommand{\LossFn}{L}
\newcommand{\Parameters}{\theta}

\newcommand{\entity}{item}
\newcommand{\entities}{items}

\DeclareMathOperator*{\argmin}{arg\,min}
\renewcommand{\log}[1]{\text{log}{\left(#1\right)}}
\renewcommand{\exp}[1]{e^{#1}}

\newcommand{\RankModel}{\NeuralRankCutoff{}}

\newcommand{\Neural}{CNN}
\newcommand{\NeuralLogistic}{\Neural{}-p}
\newcommand{\NeuralRankCutoff}{\Neural{}}
\newcommand{\LeaveOutRankSVM}{RankSVM}

\newcommand{\AllTerms}{Full}
\newcommand{\TF}{TF}
\newcommand{\TFIDF}{TF-IDF}
\newcommand{\logTFIDF}{logTF-IDF}
\newcommand{\RelativeEntropy}{RE}
\newcommand{\Multinomial}{Random $k$}
\newcommand{\BernouilliProcess}{Random \%}

\newcommand{\SubjectAllTerms}{\AllTerms{}}
\newcommand{\BodyAllTerms}{\AllTerms{}}
\newcommand{\SubjectBodyAllTerms}{\AllTerms{}}

\newcommand{\SubjectTF}{\TF{}}
\newcommand{\BodyTF}{\TF{}}
\newcommand{\SubjectBodyTF}{\TF{}}

\newcommand{\SubjectTFIDF}{\TFIDF{}}
\newcommand{\BodyTFIDF}{\TFIDF{}}
\newcommand{\SubjectBodyTFIDF}{\TFIDF{}}

\newcommand{\SubjectlogTFIDF}{\logTFIDF{}}
\newcommand{\BodylogTFIDF}{\logTFIDF{}}
\newcommand{\SubjectBodylogTFIDF}{\logTFIDF{}}

\newcommand{\SubjectMultinomial}{\Multinomial{}}
\newcommand{\BodyMultinomial}{\Multinomial{}}
\newcommand{\SubjectBodyMultinomial}{\Multinomial{}}

\newcommand{\SubjectBernouilliProcess}{\BernouilliProcess{}}
\newcommand{\BodyBernouilliProcess}{\BernouilliProcess{}}
\newcommand{\SubjectBodyBernouilliProcess}{\BernouilliProcess{}}

\newcommand{\SubjectRelativeEntropy}{\RelativeEntropy{}}
\newcommand{\BodyRelativeEntropy}{\RelativeEntropy{}}
\newcommand{\SubjectBodyRelativeEntropy}{\RelativeEntropy{}}

\newcommand{\AvocadoCollection}{Avocado}
\newcommand{\InternalCollection}{PIE}

\newcommand{\ResearchQuestionOne}{Do convolutional neural networks (\Neural{}) improve ranking efficacy over state-of-the-art query formulation methods?}
\newcommand{\ResearchQuestionTwo}{When do \Neural{}s work better than non-neural methods on the attachable item recommendation task?}
\newcommand{\ResearchQuestionThree}{What features are most important when training \Neural{}s?}

\newcommand{\RecipRank}{MRR}
\newcommand{\NDCG}{NDCG}
\newcommand{\PrecisionCut}{P@5}

\begin{document}

\title{Reply With: Proactive Recommendation of Email Attachments}

{
\normalsize

\author{Christophe Van Gysel}
\authornote{Work performed while the first author was at {Microsoft}, London.}
\orcid{0000-0003-3433-7317}
\email{cvangysel@uva.nl}
\affiliation{%
  \institution{University of Amsterdam}
  \city{Amsterdam}
  \country{The Netherlands}
}

\author{Bhaskar Mitra}
\email{bmitra@microsoft.com}
\author{Matteo Venanzi}
\email{mavena@microsoft.com}
\author{Roy Rosemarin}
\email{rorosema@microsoft.com}
\affiliation{%
  \institution{Microsoft}
  \country{United Kingdom}
}

\author{Grzegorz Kukla}
\email{grkukla@microsoft.com}
\author{Piotr Grudzien}
\email{a-pigrud@microsoft.com}
\author{Nicola Cancedda}
\email{nicanced@microsoft.com}
\affiliation{%
  \institution{Microsoft}
  \country{United Kingdom}
}

}

\renewcommand{\shortauthors}{Van Gysel et al.}

\begin{abstract}
Email responses often contain items---such as a file or a hyperlink to an external document---that are attached to or included inline in the body of the message. Analysis of an enterprise email corpus reveals that 35\% of the time when users include these items as part of their response, the attachable item is already present in their inbox or sent folder. A modern email client can proactively retrieve relevant attachable items from the user's past emails based on the context of the current conversation, and recommend them for inclusion, to reduce the time and effort involved in composing the response. In this paper, we propose a weakly supervised learning framework for recommending attachable items to the user. As email search systems are commonly available, we constrain the recommendation task to formulating effective search queries from the context of the conversations. The query is submitted to an existing IR system to retrieve relevant items for attachment. We also present a novel strategy for generating labels from an email corpus---without the need for manual annotations---that can be used to train and evaluate the query formulation model. In addition, we describe a deep convolutional neural network that demonstrates satisfactory performance on this query formulation task when evaluated on the publicly available Avocado dataset and a proprietary dataset of internal emails obtained through an employee participation program.
\end{abstract}

\maketitle

\global\csname @topnum\endcsname 0
\global\csname @botnum\endcsname 0


\section{Introduction}
\label{sec:intro}

\begin{figure}[t]
\center
\center
\includegraphics[width=1.0\linewidth]{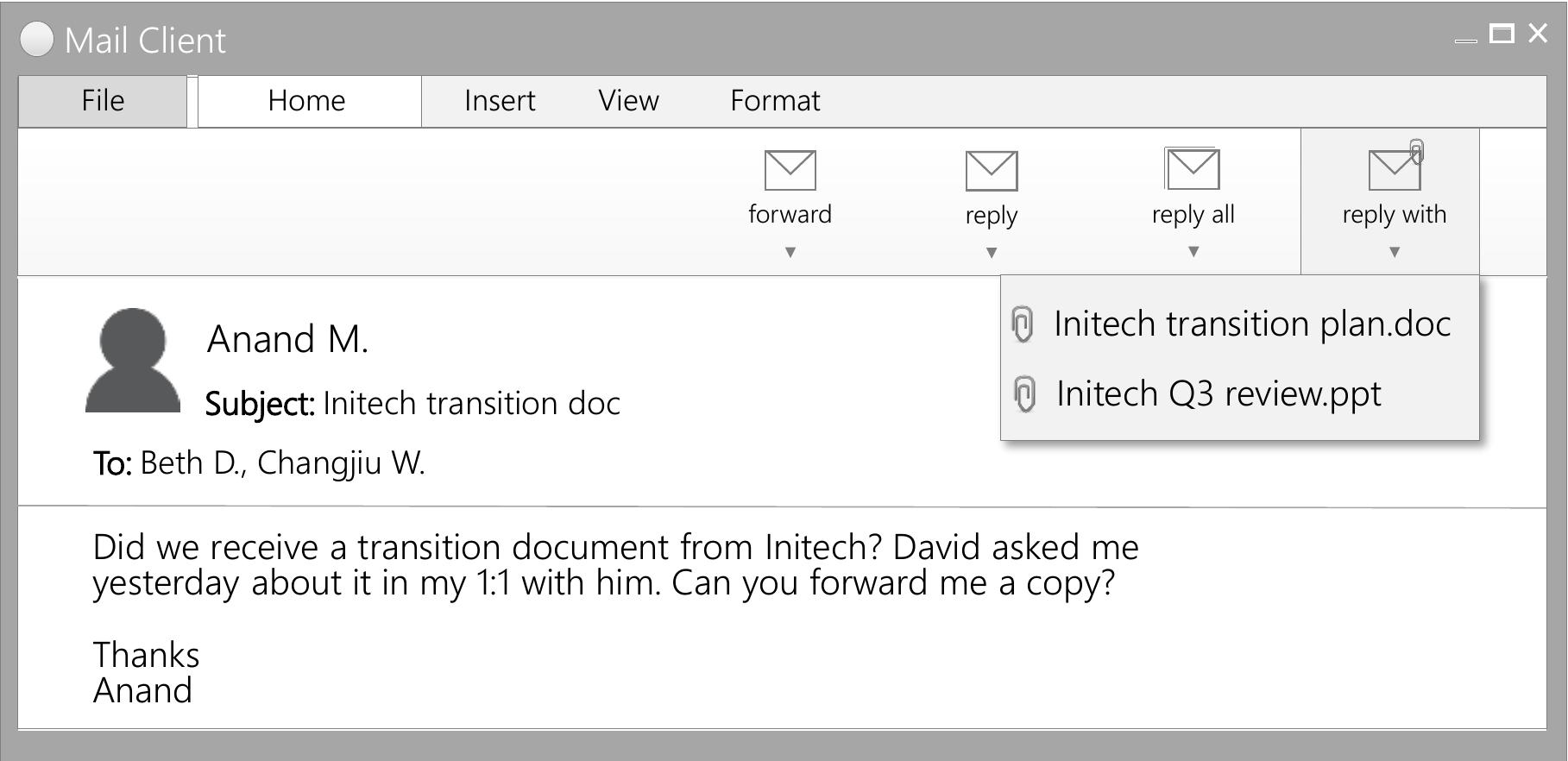}
\vspace*{-1.5\baselineskip}
\caption{Anand asks Beth and Changjiu to forward him a copy of the Initech\protect\footnotemark{} transition document. Beth's email client recommends two files, part of earlier emails in Beth's mailbox, for her to attach to her reply.}
\label{fig:example}
\end{figure}
\footnotetext{Initech is a fictional company from a popular 1990s comedy film. Any resemblance to real organizations is purely coincidental.}

In spite of the growing popularity of social networks and other modern online communication tools, email is still pervasive in the enterprise space \cite{PewResearch2014dominant}. Users typically respond to incoming emails with textual responses. However, an analysis of the publicly available Avocado dataset \cite{Oard2015avocado} reveals that 14\% of those messages also contain items, such as a file or a hyperlink to an external document. Popular email clients already detect when users forget to attach files by analyzing the text of the response message \cite{Dredze2006forgotattachment, dredze2008intelligent}. On Avocado, we find that in 35\% of the cases where the response contains attachments, the item being attached is also present in the sender's mailbox at the time of composing the response. This implies that modern email clients could help users compose their responses faster by proactively retrieving and recommending relevant items that the user may want to include with their message.

In \emph{proactive} information retrieval (IR) systems \cite{Liebling2012anticipatory, Song2016query, Benetka2017anticipating, Shokouhi2015queries}, the user does not initiate the search. Instead, retrieval is triggered automatically based on a user's current context. The context may include the time of day \cite{Song2016query}, the user's geographic location \cite{Benetka2017anticipating}, recent online activities \cite{Shokouhi2015queries} or some other criteria. In our scenario, retrieval is based on the context of the current conversation, and in particular, the message the user is responding to. In a typical IR scenario, items are ranked based on query-dependent feature representations. In the absence of an explicit search query from the user, proactive IR models may formulate a keyword-based search query using the available context information and retrieve results for the query using a standard IR model \cite{Liebling2012anticipatory}. Search functionalities are available from most commercial email providers and email search has been studied in the literature \cite{Ai2017emailsearch}. Therefore, we cast the email attachment recommendation problem as a query formulation task and use an existing IR system to retrieve emails. Attachable items are extracted from the retrieved emails and a ranking is presented to the user.

Fig.~\ref{fig:example} shows an example of an email containing an explicit request for a file. In general, there may or may not be an explicit request, but it may be appropriate to attach a relevant file with the response. Our task is to recommend the correct ``transition document'' as an attachment when Beth or Changjiu is responding to this email. In order to recommend an attachment, the model should formulate a query, such as ``\emph{Initech transition}'', based on the context of the request message, that retrieves the correct document from Beth's or Changjiu's mailbox. To formulate an effective query, the model must identify the discriminative terms in the message from Anand that are relevant to the actual file request.

Machine learning models that aim to solve the query formulation task need reliable feedback on what constitutes a good query. One option for generating labeled data for training and evaluation involves collecting manual assessments of proposed queries or individual query terms. However, it is difficult for human annotators to determine the ability of a proposed query to retrieve relevant items given only the request message. In fact, the efficacy of the query depends on the target message that should be retrieved, as well as how the IR system being employed functions. The relevance of the target message, in turn, is determined by whether they include the correct item that should be attached to the response message. Therefore, instead we propose an evaluation framework that requires an email corpus but no manual assessments. Request/response message pairs are extracted from the corpus and the model, that takes the request message as input, is evaluated based on its ability to retrieve the items attached to the response message. An IR system is employed for the message retrieval step, but is treated as a \emph{black box} in the context of evaluation. Our framework provides a concise specification for the email attachment recommendation task (\S\ref{sec:method}).

Our proposed approach for training a deep convolutional neural network (CNN) for the query formulation step is covered in \S\ref{sec:model}. The model predicts a distribution over all the terms in the request message and terms with high predicted probability are selected to form a query. Model training involves generating a dataset of request/attachment pairs similar to the case of evaluation. Candidate queries are algorithmically synthesized for each request/attachment pair such that a message from the user's mailbox with the correct item attached is ranked highly. We refer to synthetic queries as the \emph{silver-standard queries} (or silver queries for brevity) to emphasize that they achieve reasonable performance on the task, but are potentially sub-optimal. The neural model is trained to minimize the prediction loss w.r.t. the silver queries given the request message as input.

The key contributions of this paper are as follows.
\begin{inparaenum}[(1)]
	\item We introduce a novel proactive retrieval task for recommending email attachments when composing a response to an email message. The task involves formulating a query such that relevant attachable items are retrieved. Towards this goal, an evaluation framework is identified that requires an email corpus, but no manual annotations or assessments.
	\item We show how a machine learning model can be trained on an unlabeled email corpus for this query formulation task. The training involves generating a dataset of request/response message pairs and corresponding sets of silver queries.
	\item Finally, we present a neural model for estimating a probability distribution over the terms in the request message for formulating queries.
\end{inparaenum}


\section{Related Work}
\label{sec:related_work}

\subsection{Proactive IR}

\emph{Zero-query} search---or proactive IR---scenarios have received increasing attention recently \cite{Allan2012frontiers}. However, similar approaches have also been studied in the past under other names, such as \emph{just-in-time} \cite{Rhodes2000just, Rhodes1996remembrance, Rhodes2000margin, Rhodes1997wearable}, query-free \cite{Hart1997query} or anticipatory \cite{Liebling2012anticipatory, Budzik1999watson} IR. According to \citet{Hart1997query}, the goal of the proactive retrieval system is to surface information that helps the user in a broader task. While some of these works focus on displaying contextually relevant information next to Web pages \cite{Rhodes1996remembrance, Rhodes2000margin, Crabtree1998adaptive, Budzik1999watson, Maglio2000suitor} or multimedia \citep{Odijk2015dynamic}, others use audio cues \cite{Mynatt1998designing, Sawhney2000nomadic} or signals from other sensors \cite{Ryan1999enhanced, Rhodes1997wearable} to trigger the retrieval.
In more recent years, proactive IR systems have re-emerged in the form of intelligent assistant applications on mobile devices, such as Siri, Google Now and Cortana. The retrieval in these systems may involve modeling repetitive usage patterns to proactively show concise information cards \cite{Shokouhi2015queries, Song2016query} or surface them in response to change in user context such as location \cite{Benetka2017anticipating}.
\citet{Hart1997query}, \citet{Budzik1999watson} and \citet{Liebling2012anticipatory} propose to proactively formulate a query based on user's predicted information need.
In contrast to previous work on proactive contextual recommendation, we formulate a query to retrieve attachable items to assist users with composing emails instead of supplying information to support content or triggering information cards in mobile assistants.

\subsection{Predictive models for email}

Email overload is the inability to effectively manage communication due to the large quantity of incoming messages \citep{Whittaker1996overload}. \citet{Grevet2014overload} find that work email tends to be overloaded due to outstanding tasks or reference emails saved for future use. \citet{Ai2017emailsearch} find that \numprint{85}\% of email searches are targeted at retrieving known items (e.g., reference emails, attachments) in mailboxes. \citet{Horvitz1999principles} argues that a combination of two approaches---%
\begin{inparaenum}[(1)]
	\item providing the user with powerful tools, and
	\item predicting the user's next activity and taking automated actions on her behalf
\end{inparaenum}%
---is effective in many scenarios. Modern email clients may better alleviate email overload and improve user experience by incorporating predictive models that try to anticipate the user's need and act on their behalf.

Missing attachments in email generates a wave of responses notifying the sender of her error. \citet{Dredze2006forgotattachment} present a method that notifies the user when a file should be attached before the email is sent. \citet{DiCastro2016youvegotmail} proposed a learning framework to predict the action that will be performed on an email by the user, with the aim of prioritizing actionable emails. \citet{Carvalho2005collective} classify emails according to their speech acts. \citet{Graus2014recipientrecommendation,Qadir2016activity} recommend recipients to send an email message to. \citet{Kannan2016smart} propose an end-to-end method for automatically generating email responses that can be sent by the user with a single click.

\subsection{Query formulation and reformulation}

The task of contextual recommendation of attachable \entities{} by means of query formulation has not received much attention. However, there is work on query extraction from verbose queries and query construction for related patent search. Similar to this paper, the methods below consider the search engine as a black box.

\paragraph{Prior art search}

Establishing novelty is an important part of the patenting process. Patent practitioners (e.g., lawyers, patent office examiners) employ a search strategy where they construct a query based on a new patent application in order to find prior art. However, patents are different from typical documents due to their length and lack of mid-frequency terms \citep{Lupu2013patentsurvey}.

Automated query generation methods have been designed to help practitioners search for prior art. \citet{Xue2009transformingpatents} use TF-IDF to generate a ranking of candidate query terms, considering different patent fields, to rank similar patents. In later work \citep{Xue2009querygenerationpatents}, they incorporated a feature combination approach to further improve prior art retrieval performance. Alternative term ranking features, such as relative entropy \citep{Mahdabi2011buildingqueriespriorart}, term frequency and log TF-IDF \citep{Cetintas2012querygenerationpriorart}, have also been explored. \citet{Kim2011automaticbooleanquerysuggestion} suggest boolean queries by extracting terms from a pseudo-relevant document set. \citet{Golestan2015patent} find that an interactive relevance feedback approach outperforms state-of-the-art automated methods in prior art search.

Query extraction methods used for prior art search can also be applied to the task of attachable \entity{} recommendation considered in this paper. Consequently, we consider the methods mentioned above as our baselines (\S\ref{sec:baselines}). However, there are a few notable differences between the patent and email domains:
\begin{inparaenum}[(1)]
	\item Email messages are much shorter in length than patents.
	\item Patents are more structured (e.g., US patents contain more than 50 fields) than email messages.
	\item Patents are linked together by a static citation graph that grows slowly, whereas email messages are linked by means of a dynamic conversation that is fast-paced and transient in nature.
	\item In the case of email, there is a social graph between email users that can act as an additional source of information.
\end{inparaenum}

\paragraph{Improving verbose queries}

\citet{Bendersky2008discoveringkeyconcepts} point out that search engines do not perform well with verbose queries \citep{Balasubramanian2010webqueryreductions}. \citet{Kumaran2009reducinglongqueries} propose a sub-query extraction method that obtains ground truth by considering every sub-query of a verbose query and cast it as a learning to rank problem. \citet{Xue2010verbosequeries} use a Conditional Random Field (CRF) to predict whether a term should be included. However, inference using their method becomes intractable in the case of long queries. \citet{Lee2009querytermsranking} learn to rank query terms instead of sub-queries with a focus on term dependency. \citet{Huston2010evaluatingverbosequeryprocessing} find that removing the stop structure in collaborative question answering queries increases retrieval performance. \citet{Maxwell2013compactquerytermselection} propose a method that selects query terms based on a pseudo-relevance feedback document set. \citet{Meij2009learningsemanticquery} identify semantic concepts within queries to suggest query alternatives. Related to the task of improving verbose queries is the identification of important terms \citep{Zhao2010termnecessity}. \citet{He2004scs} note that the use of relevance scores for query performance prediction is expensive to compute and focus on a set of pre-retrieval features that are strong predictors of the query's ability to retrieve relevant documents.{} See \cite{Gupta2015verboseyqueriessurvey} for an overview.

The task we consider in this paper differs from search sub-query selection as follows.
\begin{inparaenum}[(1)]
	\item Search queries are formulated by users as a way to interface with a search engine. Requests in emails may be more complex as they are formulated to retrieve information from a human recipient, rather than an automated search engine. In other words, email requests are more likely to contain natural language and figurative speech than search engine queries. This is because the sender of the request does not expect their message to be parsed by an automated system.
	\item Search sub-query extraction aims to improve retrieval effectiveness while the query intent remains fixed. This is not necessarily the case in our task, as a request message like has the intent to retrieve information from the recipient (rather than a retrieval system operating on top of the recipient's mailbox).
	\item Work on search sub-query selection \citep{Kumaran2009reducinglongqueries,Xue2010verbosequeries} takes advantage of the fact that 99.9\% of search queries consist of 12 terms or less \citep{Bendersky2009analysis} by relying on computations that are intractable otherwise. As emails are longer (Table~\ref{tbl:statistics}), many of the methods designed for search sub-query selection are not applicable in our setting.
\end{inparaenum}


\section{Proactive attachable item recommendation}
\label{sec:method}

\begin{figure}[t!]
\centering
\includegraphics[width=\linewidth]{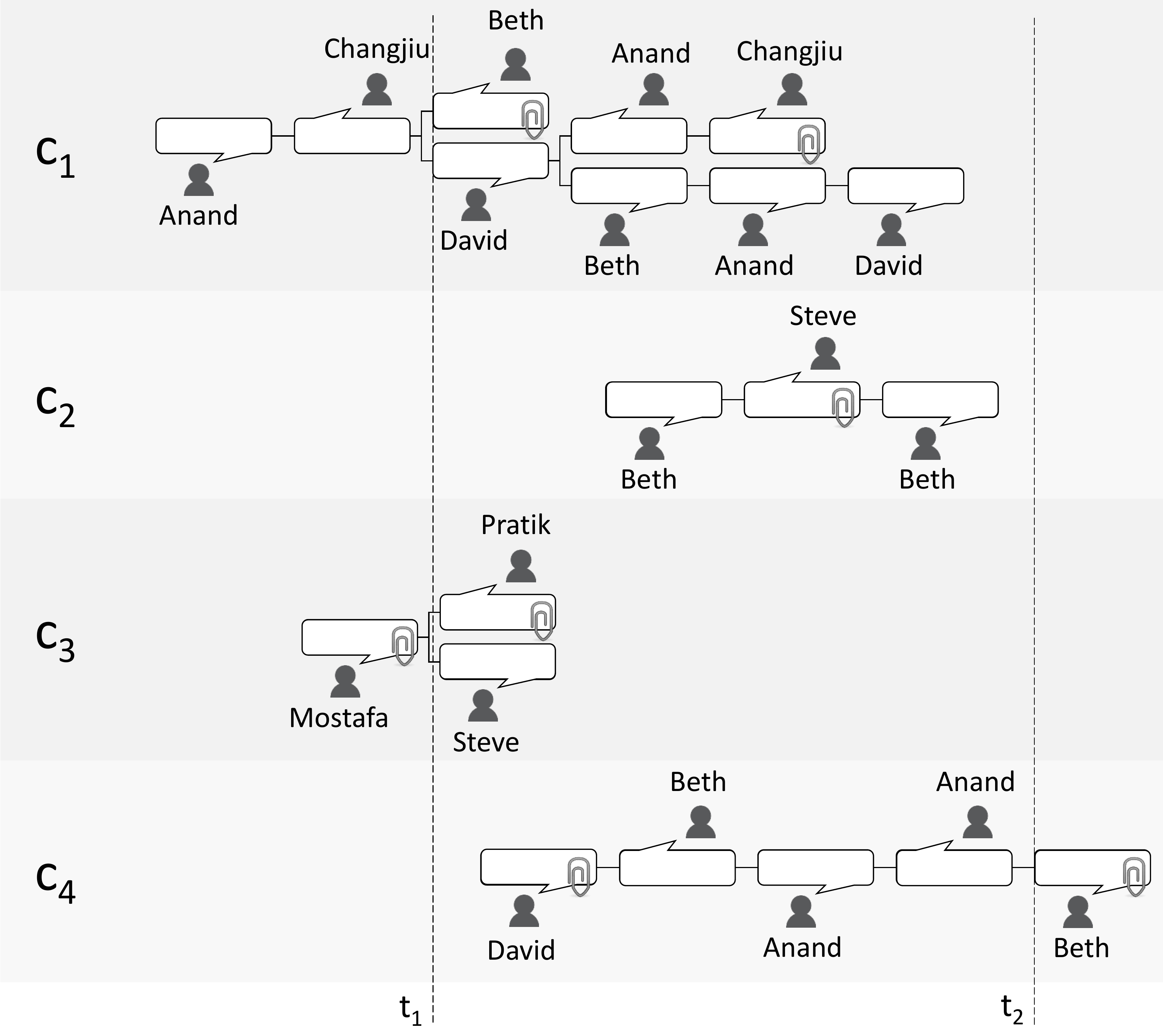}
\vspace*{-2\baselineskip}
\caption{Beth's mailbox has four on-going conversations. When Beth replies with an attachment in conversation $\Thread{}_{1}$, at time $\Time{}_{1}$, only the attachment she received from Mostafa (conversation $\Thread{}_{3}$) is present in her mailbox.}
\label{fig:framework}
\end{figure}

\newcommand{\FirstUser}{\User{}_{a}}
\newcommand{\SecondUser}{\User{}_{b}}

\newcommand{\ExampleRequest}{\Message{}_{a\rightarrow b}}
\newcommand{\ExampleResponse}{\Message{}_{b\rightarrow a}}

\newcommand{\ExampleRequestReplyPair}{\left\langle\Message_{a\rightarrow b}, \Message_{b\rightarrow a}\right\rangle}

Given message $\ExampleRequest{}$ from user $\FirstUser{}$ to user $\SecondUser{}$, we want to recommend an item $\AttachableEntity{}$ that the receiver $\User{}_{b}$ may want to attach (or include) in the response $\ExampleResponse$. Email corpora, such as Avocado \cite{Oard2015avocado}, contain many conversation threads where each conversation $\Thread{}$ contains messages $\Message_{i} \in \Thread{}$ exchanged between several participants. From these conversations, we can identify pairs of request-response messages $\ExampleRequestReplyPair{}$ where $\ExampleResponse{}$ contains an attachment $\TargetAttachableEntity{}$. We assume that user $\SecondUser{}$ included $\TargetAttachableEntity{}$ in $\ExampleResponse{}$ in response to an explicit or an implicit request in the message $\ExampleRequest{}$. Such pairs of request message and attachment $\ExampleRequestReplyPair{}$ form the ground-truth in our evaluation framework.

\newcommand{\BethUser}{\User{}_{\text{Beth}}}
\newcommand{\DavidUser}{\User{}_{\text{David}}}
\newcommand{\ChangjiuUser}{\User{}_{\text{Changjiu}}}
\newcommand{\MostafaUser}{\User{}_{\text{Mostafa}}}
\newcommand{\SteveUser}{\User{}_{\text{Steve}}}

Fig.~\ref{fig:framework} shows a sample mailbox $\Mailbox{}_{\text{Beth}}$ of user $\BethUser{}$ containing four conversations $\{\Thread{}_{1}, \Thread{}_{2}, \Thread{}_{3}, \Thread{}_{4}\}$. During these conversations, $\BethUser{}$ responds with an attachment twice---at time $\Time{}_{1}$ and $\Time{}_{2}$. At time $\Time{}_{1}$, in this toy example, the set of candidate items that are available in the user's mailbox for recommendation contains only the attachment from $\MostafaUser{}$ received during the conversation $\Thread{}_{3}$. At $\Time{}_{2}$, however, the set of candidates includes attachments received on all four conversation threads---from $\ChangjiuUser{}$ ($\Thread{}_{1}$), $\SteveUser{}$ ($\Thread{}_{2}$), $\MostafaUser{}$ ($\Thread{}_{3}$), $\User{}_{\text{Pratik}}$ ($\Thread{}_{3}$), and $\DavidUser{}$ ($\Thread{}_{4}$)---as well as the item sent by $\BethUser{}$ previously on the conversation thread $\Thread{}_{1}$.

It is important to emphasize that our problem setting has two important constraints when recommending items, that any model should adhere to
\begin{inparaenum}[(1)]
	\item a \textbf{privacy} constraint: the model can only recommend items from a user's own mailbox, and
	\item a \textbf{temporal} constraint: the model can only recommend items that are already present in the user's mailbox at the time of recommendation.
\end{inparaenum}

\subsection{Attachment retrieval}

In addition to the above domain-specific constraints, we limit our setup to using a standard IR system $\RetrievalSystem{}$ for retrieval, and cast the problem that the model needs to solve as a query formulation task. Using an existing IR system has the practical benefit that one only needs to maintain a single system in contrast to the alternative where a separate attachment recommendation engine needs to be maintained. The model is presented with a message $\RequestMessage{}$ containing an explicit or an implicit content request. The model is tasked with generating a query that can be submitted to the retrieval system $\RetrievalSystem{}$ that retrieves a set of ranked messages $\Messages{}_{\RetrievalSystem{}}$ from the user's mailbox. Under this assumption, the retrieval system $\RetrievalSystem{}$ is treated as a black box, and we are only interested in optimizing the query formulation model. Note that a query is only formulated when it is clear that an item needs to be attached to a reply message (\S\ref{sec:rq}), such as is the topic of \citep{Dredze2006forgotattachment,dredze2008intelligent}.

To extract a ranked list of attachable items from search engine $\RetrievalSystem{}$, we adopt an approach popular in entity retrieval frameworks \citep{Balog2006expertfinding} where an entity model is the mixture of document models that the entity is associated with. For a given query $\Query{}$ issued at time $\CurrentTime{}$ in the mailbox of user $\User{}$, attachable items $\AttachableEntity{} \in \AttachableEntities{}$ are then ranked in decreasing order of
\newcommand{\RelevanceProb}{\CondProb{\AttachableEntity{}}{\Query{}, \User{}, \CurrentTime{}}}%
\newcommand{\RetrievalSystemProb}{\Prob[S_\RetrievalSystem{}]}%
\newcommand{\AssociationStrength}{\Apply{f}{\AttachableEntity{} \mid \Message{}}}
\begin{equation}
\label{eq:ranking}
\RelevanceProb{} \propto \frac{1}{\Apply{Z_1}{\AttachableEntity{}, \User{}, \CurrentTime{}}} \sum_{\substack{\Message{} \in \Mailbox{} \\ \MessageTime{} < \CurrentTime{}}} \CondProb[\RetrievalSystemProb]{\Message{}}{\Query{}} \AssociationStrength{}
\end{equation}
where $\CondProb[\RetrievalSystemProb]{\Message{}}{\Query{}}$ is the relevance score for message $\Message{}$ given query $\Query{}$ according to retrieval system $\RetrievalSystem{}$, $\MessageTime{}$ is the timestamp when the message $\Message{}$ appeared first in the mailbox $\Mailbox{}$, $\CurrentTime{}$ is time when the model needs to make the recommendation, $\AssociationStrength{}$ denotes the association strength between message $\Message{}$ and \entity{} $\AttachableEntity{}$, and
\newcommand{\NormalizationConstant}{\Apply{Z_1}{\AttachableEntity{}, \User{}, \CurrentTime{}}}%
$
\NormalizationConstant{} = \sum_{\substack{\Message{} \in \Mailbox{} \\ \MessageTime{} < \CurrentTime{}}} \AssociationStrength{}
$
is a normalization constant. The normalization constant $\NormalizationConstant$ avoids a bias towards attachable items that are associated with many messages (e.g., electronic business cards).

We associate messages with an attachable item according to the presence of the item within a message and its surrounding messages within the conversation:
$
\AssociationStrength{} = \Apply{\mathds{1}_{\Apply{\text{context}}{\Message{}}}}{\AttachableEntity{}}.
$
In this paper, we take $\Apply{\text{context}}{\Message{}}$ to be all messages $\Message{}^\prime$ in the same conversation $\Thread{}_\Message{}$ as message $\Message{}$ that occurred before the time of recommendation, i.e., $\Time{}_{\Message{}^\prime} < \CurrentTime{}$. Note that the exact definition of an attachable \entity{} depends on the email domain and can include individual files, file bundles and hyperlinks to external documents amongst others (see \S\ref{sec:collections}).

\subsection{Evaluating query formulations}
\label{sec:framework-eval}

Once we have extracted $\RequestEntityPair{}$ pairs from an email corpus, each request message $\RequestMessage{}$ is presented to the query formulation model that we want to evaluate. The model generates a query $\Query{}$ conditioned on the message $\RequestMessage{}$. The query $\Query{}$ is submitted to retrieval system $\RetrievalSystem{}$ and attachable items extracted from the retrieved messages are determined according to Eq.~\ref{eq:ranking}. Given the ranked list of attachable \entities{} $\AttachableEntities{}_{\text{retrieved}}$ and the expected item $\TargetAttachableEntity{}$ we can compute standard rank-based IR metrics such as MRR and NDCG (\S\ref{sec:metrics}). We report the mean metric over all $\RequestEntityPair{}$ pairs extracted from the corpus. 

Our approach of using the historical information from an email corpus for evaluation is comparable to the application of \emph{click-through} data for similar purposes in Web search.
In the document retrieval scenario, a user's click on a document $\Document{}$ on the search result page is considered an implicit vote of confidence on the relevance of $\Document{}$ to the query. Learning to rank models can be trained on this \emph{click-through} data \cite{Joachims2002optimizing, xu2010improving, macdonald2009usefulness} if human relevance judgments are not available in adequate quantity. By explicitly attaching a file $\TargetAttachableEntity{}$, similarly, the user of an email system provides a strong indication that recommending $\TargetAttachableEntity{}$ at the time of composing $\ReplyMessage{}$ would have been useful. We can use this information to train and evaluate supervised models for ranking attachments at the time of email composition.


\section{Query formulation model}
\label{sec:model}

We first introduce a method for generating pseudo training data \citep{Asadi2011pseudo,Berendsen2013pseudo} without the need for manual assessments. Silver-standard queries are algorithmically synthesized for each request/attachment pair and consequently scored by measuring the query's ability to retrieve the relevant attachment (\S\ref{sec:model:subquery-generation}). The request/query pairs part of the pseudo training collection are then used to train a convolutional neural network (\S\ref{sec:model:neural}) that learns to extract query terms from a request message. In this paper, we use a convolutional architecture rather than a recurrent one, as we intend to model term importance by term context without relying on the exact ordering of terms.

\subsection{Model training}
\label{sec:model:subquery-generation}

The candidate silver queries are extracted for request-response pairs $\RequestReplyPair{}$ in a training set. Given a request message $\RequestMessage{}$ and its associated target attachable \entity{} $\TargetAttachableEntity{} \in \AttachableEntities{}_{\ReplyMessage{}}$ that is attached to reply $\ReplyMessage{}$, where $\AttachableEntities{}_{\ReplyMessage{}}$ is the set of \entities{} attached to $\ReplyMessage{}$, the objective is to select the $\CandidateQueryTermBudget{}$ terms that are most likely to retrieve \entity{} $\TargetAttachableEntity{}$ according to Eq.~\ref{eq:ranking}.

In the ideal case, one considers the powerset of all terms within request message $\RequestMessage{}$ as candidate silver queries \citep{Kumaran2009reducinglongqueries,Xue2010verbosequeries}. However, considering all terms is computationally intractable in our case as email messages tend to average between \numprint{70} to \numprint{110} tokens (Table~\ref{tbl:statistics}).

In order to circumvent the intractability accompanied with computing the powerset of all terms in a message, we use the following stochastic strategy to select a fixed number of candidate query terms that we compute the powerset of. We consider two sources of query terms.
\newcommand{\RecallableMessagesFn}{\Messages{}_{\text{recallable}}}
\newcommand{\RecallableMessages}{\Apply{\RecallableMessagesFn{}}{\TargetAttachableEntity{}, \CurrentTime{}}}
The first source of candidate query terms consists of \textbf{subject} terms: topic terms in the subject of the request message. Email subjects convey relevance and context \citep{Weil2004overcomingoverload} and can be seen as a topical summary of the message. For the second source of query terms, we consider \textbf{recallable} terms: infrequent terms that occur frequently in messages $\RecallableMessages{} = \{\Message{} \in \Mailbox{} \mid \MessageTime{} < \CurrentTime{}, \TargetAttachableEntity{} \in \AttachableEntities{}_\Message{} \}$ that contained \entity{} $\TargetAttachableEntity{}$ and occur at least once in the request message. That is, we gather all terms that have the potential to retrieve $\TargetAttachableEntity{}$ (according to Eq.~\ref{eq:ranking}) and select those terms that occur in at least 30\% of messages $\RecallableMessages{}$ and occur in less than 1\% of all messages.

\begin{algorithm}[t!]
\newcommand{\CandidateTerms}{T}
\newcommand{\CandidateTerm}{t}

\SetAlgoLined
\SetNoFillComment
\setstretch{0.45}

\SetKwFunction{isUnwanted}{isUnwanted}

\KwData{request message $\RequestMessage{}$, query term budget $\CandidateQueryTermBudget{}$}
\KwResult{candidate silver queries $\RequestMessageCandidateQueries{}$}
set of candidate terms $\CandidateTerms{} \gets$ \{\}\;
\While{$($term candidates left $\wedge$ $\Length{\CandidateTerms{}} < \CandidateQueryTermBudget{}$$)$}{
	$S \gets$ uniformly random choose \textbf{subject} or \textbf{recallable terms}\;
	\If{$S = \varnothing$}{
		\Continue
	}
	$\CandidateTerm{} \gets \argmin_{\CandidateTerm{} \in S} \Apply{\text{df}}{\CandidateTerm{}}$\;
	\If{$\neg$\isUnwanted{$\CandidateTerm{}$}}{
		$\CandidateTerms{} = \CandidateTerms{} \, \cup \, \{t\}$\;
	}
	$S \gets S \setminus \{\CandidateTerm{}\}$
}
$\RequestMessageCandidateQueries \gets 2^{\CandidateTerms{}} - \{\varnothing\}$
\caption{Candidate query terms are selected by choosing a random query term source and selecting the query term with the lowest document frequency while ignoring unwanted query terms \citep{Salton1983boolean}. The \texttt{isUnwanted} predicate is true when the term is a stopword, contains a digit, contains punctuation or equals the names of the email sender/recipients. After selecting $\CandidateQueryTermBudget{}$ terms, we consider the powerset of selected terms as silver queries.\label{alg:subquery-generation}}
\end{algorithm}

To construct candidate silver queries for a request message $\RequestMessage{}$, we follow the strategy as outlined in Algorithm~\ref{alg:subquery-generation} that mimics the boolean query formulation process of \citet{Salton1983boolean}. Candidate terms are selected from either the \textbf{subject} or \textbf{recallable} source in increasing order of document frequency (i.e., infrequent terms first). Unwanted terms, such as stopwords, digits, punctuation and the names of the sender and recipients that occur in the email headers, are removed. Afterwards, we take the candidate queries $\RequestMessageCandidateQueries{}$ to be all possible subsets of candidate terms (excluding the empty set).

Once we have obtained the set of candidate queries $\RequestMessageCandidateQueries{}$ for request message $\RequestMessage{}$ we score the candidate queries as follows. For every $\CandidateQuery{} \in \RequestMessageCandidateQueries{}$ we rank email messages using retrieval system $\RetrievalSystem{}$ according to $\CandidateQuery{}$. We then apply Eq.~\ref{eq:ranking} to obtain a ranking over \entities{} $\UserTimeConstraintedAttachableEntities{}$ in the mailbox of user $\User{}$ at time $\CurrentTime{}$. As we know the target \entity{} $\TargetAttachableEntity{}$ to be retrieved for request message $\RequestMessage{}$, we quantify the performance of candidate query $\CandidateQuery{}$ by its reciprocal rank,
\newcommand{\CandidateQueryScore}[1]{\Apply{\text{score}}{#1}}
$
\CandidateQueryScore{\CandidateQuery{}} = \frac{1}{\Apply{\text{rank}}{\TargetAttachableEntity{}}} \in (0, 1]
$
where $\Apply{\text{rank}}{\TargetAttachableEntity{}}$ denotes the position of \entity{} $\TargetAttachableEntity{}$ (Eq.~\ref{eq:ranking}) in the item ranking.

After computing the score for every candidate silver query, we group queries that perform at the same level (i.e., that have the same score) for a particular request message $\RequestMessage{}$. We then apply two post-processing steps that improve silver-standard query quality based on the trade-off between query broadness and specificness. Following \citet{Salton1983boolean} on boolean query formulation, specific queries are preferred over broad queries to avoid loss in precision. Queries can be made more specific by adding terms. Consequently, within every group of equally-performing queries, we remove subset queries whose union results in another query that performs at the same level as the subsets. For example, if the queries \emph{``barack obama''}, \emph{``obama family''} and \emph{``barack obama family''} all achieve the same reciprocal rank, then we only consider the latter three-term query and discard the two shorter, broader queries. An additional argument for the strategy above follows from the observation that any term not part of the query is considered as undesirable during learning. Therefore, including all queries listed above as training material would introduce a negative bias against the terms \emph{``barack''} and \emph{``family''}. However, queries that are too specific can reduce the result set \citep{Salton1983boolean} or cause query drift \citep{Mitra1998drift}. Therefore, the second post-processing step constitutes the removal of supersets of queries that perform equal or worse. The intuition behind this is that the inclusion of the additional terms in the superset query did not improve retrieval performance. For example, if queries \emph{``barack obama''} and \emph{``barack obama president''} perform equally well, then the addition of the term \emph{``president''} had no positive impact on retrieval. Consequently, including the superset query (i.e., \emph{``barack obama president''}) in the training set is likely to motivate the inclusion of superfluous terms that negatively impact retrieval effectiveness.

\subsection{A convolutional neural network for ranking query terms}
\label{sec:model:neural}

\newcommand{\circled}[2][yscale=0.6]{%
\begin{tikzpicture}[#1, xscale=1.0, baseline=(char.base)]%
\node[shape=ellipse, draw, inner sep=1.5pt, text width=0.35cm] (char) {#2};
\end{tikzpicture}%
}

\newcommand{\EndOfRanking}[1]{\scalebox{1}[0.8]{\circled{\footnotesize{}\normalfont{}\hspace*{-0.5pt}EoR}}}
\newcommand{\CaptionEndOfRanking}{\EndOfRanking}

\begin{figure}[t]
\small
\begin{tikzpicture}[yscale=-0.30]
\node[anchor=west, align=center, text width=1cm] at (0.0, 0) {\textbf{Rank}};
\node[anchor=west, align=center, text width=4cm] at (1.0, 0) {\textbf{Term}};
\node[anchor=west, align=center, text width=1cm] at (5.2, 0) {\textbf{Score}};
\newcommand{\RankedItem}[3]{
	\node[anchor=west, align=center, text width=1cm] at (0.0, #1) {\texttt{#1.}};
	\node[anchor=west, align=center, text width=4cm] at (1.0, #1) {\texttt{#2}};
	\node[anchor=west, align=center, text width=1cm] at (5.2, #1) {\texttt{#3}};
}
\RankedItem{1}{initech}{0.20}
\RankedItem{2}{initech}{0.18}
\RankedItem{3}{transition}{0.15}
\RankedItem{4}{- - - - - -\EndOfRanking{}- - - - - -}{0.10}
\RankedItem{5}{david}{0.07}
\RankedItem{6}{...}{}
\end{tikzpicture}
\vspace*{-\baselineskip}
\caption{Terms in the request message of Fig.~\ref{fig:example} are ranked by our model. In addition to the terms, the model also ranks a \protect\CaptionEndOfRanking{} token that specifies the query end. The final query becomes \emph{``initech transition''} as duplicate terms are ignored.\label{fig:example_ranking}}

\end{figure}

After obtaining a set of candidate queries $\RequestMessageCandidateQueries{}$ for every request/\entity{} pair $\RequestEntityPair{}$ in the training set, we learn to select query terms from email threads using a convolutional neural network model that convolves over the terms contained in the email thread. Every term is characterized by its context and term importance features that have been used to formulate queries in previous work \citep{Bendersky2008discoveringkeyconcepts,Cetintas2012querygenerationpriorart,Xue2009querygenerationpatents,Mahdabi2011buildingqueriespriorart,Zhao2008sqs,Kumaran2009reducinglongqueries,He2004scs}. Our model jointly learns to
\begin{inparaenum}[(1)]
	\item generate a ranking of message terms, and
	\item determine how many terms of the message term ranking should be included in the query.
\end{inparaenum}
In order to determine the number of terms included in the query, the model learns to rank an end-of-ranking token \EndOfRanking{} in addition to the message terms. Fig.~\ref{fig:example_ranking} shows a term ranking for the example in Fig.~\ref{fig:example}. Terms in the request message are ranked in decreasing order of the score predicted by our model. Terms appearing at a lower rank than the \EndOfRanking{} are not included in the query.

\newcommand{\WeightMatrix}[1]{W_{#1}}
\newcommand{\TermEmbeddingMatrix}{\WeightMatrix{\text{repr}}}

\newcommand{\ReLU}{
	\begin{tikzpicture}
		\draw (0, 0) -- (0.5, 0);
		\draw (0.3, 0) -- (0.60, 0.50);
	\end{tikzpicture}
}

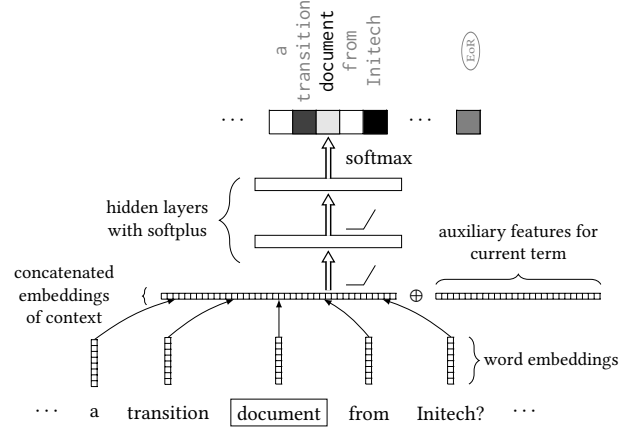
\begin{figure}[t]

\resizebox{\columnwidth}{!}{%
\begin{tikzpicture}[font=\large]

\tikzstyle{vecArrow} = [thick, decoration={markings,mark=at position
   1 with {\arrow[semithick]{open triangle 60}}},
   double distance=1.4pt, shorten >= 5.5pt,
   preaction = {decorate},
   postaction = {draw,line width=1.4pt, white,shorten >= 4.5pt}]
]

%
%

\node [rectangle, minimum height = 0.09cm, minimum width = 4cm, inner sep = 0] (concat_context) {};
\draw [anchor = bottom left, step=0.10cm, black, shift={(concat_context.south west)}] (concat_context.south west) grid (concat_context.north east);

\newcommand{\Input}[3]{
\node [anchor=base, align=center, #3] (#1) {#2};
}
\newcommand{\Word}[4]{
\Input{#1}{#2}{#3}
\node [rectangle, above = 0.25cm of #1, minimum height = 0.79cm, minimum width = 0.10cm, inner sep = 0] (#1_repr) {};
\draw [anchor = bottom left, step=0.10cm, black, shift={(#1_repr.south west)}] (#1_repr.south west) grid (#1_repr.north east);

\draw [->] (#1_repr.north) to #4;
}

\Word{document}{document}{draw, below = 1.70cm of concat_context}{(concat_context.south)}

\Word{transition}{transition}{left = 0.25 of document}{[bend left=10] ($ (concat_context.south west) + (1.25, 0) $)}
\Word{a}{a}{left = 0.25 of transition}{[bend left=10] ($ (concat_context.south west) + (0.25, 0) $)}
\Input{ellipsis_begin}{$\cdots$}{left = 0.25cm of a}

\Word{from}{from}{right = 0.25 of document}{[bend right=10] ($ (concat_context.south east) - (1.25, 0) $)}
\Word{initech}{Initech?}{right = 0.25 of from}{[bend right=10] ($ (concat_context.south east) - (0.25, 0) $)}
\Input{ellipsis_end}{$\cdots$}{right = 0.25 of initech}

\node [right = 0.1cm of concat_context] (concat_ellipsis) {$\oplus$};

\node [rectangle, right = 0.1cm of concat_ellipsis, minimum height = 0.09cm, minimum width = 2.8cm, inner sep = 0] (auxiliary) {};
\draw [anchor = bottom left, step=0.10cm, black, shift={(auxiliary.south west)}] (auxiliary.south west) grid (auxiliary.north east);

%
%

\node [rectangle, draw, above = 0.60cm of concat_ellipsis, minimum width=2.5cm, minimum height=0.2cm, xshift=-1.5cm] (hidden_first) {};
\draw[vecArrow] ($ (concat_ellipsis) + (-1.5, 0.10) $) to node [right, xshift=5pt] {\ReLU{}} ($ (hidden_first) - (0, 0.15) $);

\node [rectangle, draw, above = 0.75cm of hidden_first, minimum width=2.5cm, minimum height=0.2cm] (hidden_second) {};
\draw[vecArrow] ($ (hidden_first) + (0, 0.10) $) to node [right, xshift=5pt] {\ReLU{}} ($ (hidden_second) - (0, 0.15) $);

\node [rectangle, above = 0.75cm of hidden_second, minimum height = 0.40cm, minimum width = 2.0cm, inner sep = 0] (softmax) {};
\draw [anchor = bottom left, step=0.40cm, black, shift={(softmax.south west)}] (softmax.south west) grid (softmax.north east);

\node [anchor=base, align=center, left = 0.25cm of softmax] {$\cdots{}$};
\node [anchor=base, align=center, right = 0.25cm of softmax] (softmax_right_ellipsis) {$\cdots{}$};

\draw[vecArrow] ($ (hidden_second) + (0, 0.10) $) to node [right, xshift=5pt] {softmax} ($ (softmax) - (0, 0.25) $);

\node [rectangle, right = 0.25cm of softmax_right_ellipsis, minimum height = 0.40cm, minimum width = 0.40cm, inner sep = 0] (softmax_eor) {};
\draw [anchor = bottom left, step=0.40cm, black, shift={(softmax_eor.south west)}] (softmax_eor.south west) grid (softmax_eor.north east);

\newcommand{\SoftmaxLabel}[2]{
\node [anchor=base, align=center, above = of softmax, rotate = 90, #2] {\texttt{#1}};
}

\SoftmaxLabel{a}{yshift=0.60cm, text=gray}
\SoftmaxLabel{transition}{yshift=0.20cm, text=gray}
\SoftmaxLabel{document}{yshift=-0.20cm}
\SoftmaxLabel{from}{yshift=-0.60cm, text=gray}
\SoftmaxLabel{Initech}{yshift=-1.0cm, text=gray}

\SoftmaxLabel{\EndOfRanking{}}{yshift=-2.7cm, text=gray}

\draw [fill=black] ($ (softmax.south east) + (0, 0) $) rectangle ($ (softmax.south east) + (-0.40cm, 0.40cm) $);
\draw [fill=black!75] ($ (softmax.south west) + (0.40cm, 0) $) rectangle ($ (softmax.south west) + (0.80cm, 0.40cm) $);
\draw [fill=black!50] ($ (softmax.south east) + (1.18cm, 0) $) rectangle ($ (softmax.south east) + (1.58cm, 0.40cm) $);
\draw [fill=black!10] ($ (softmax.south west) + (0.80cm, 0) $) rectangle ($ (softmax.south west) + (1.20cm, 0.40cm) $);

%
%
\draw [decorate, decoration={brace,amplitude=10pt}, xshift=-4pt, yshift=0pt] ($ (hidden_first.south west) - (0.25, 0.25) $) -- ($ (hidden_second.north west) - (0.25, 0) $) node [black, midway, xshift=-40pt, rotate=0] {\normalsize \begin{tabular}{c}hidden layers\\with softplus\end{tabular}};

\draw [decorate, decoration={brace,amplitude=5pt,mirror}, xshift=40pt, yshift=0pt] ($ (initech_repr.south east) + (0.25, 0) $) -- ($ (initech_repr.north east) + (0.25, 0) $) node [black, midway, xshift=40pt, rotate=0] {\normalsize \begin{tabular}{c}word embeddings\end{tabular}};

\draw [decorate, decoration={brace,amplitude=2pt}, xshift=-4pt, yshift=0pt] ($ (concat_context.north west) + (-0.25, -0.20) $) -- ($ (concat_context.south west) + (-0.25, 0.20) $) node [black, midway, xshift=-40pt, rotate=0] {\normalsize \begin{tabular}{c}concatenated\\embeddings\\of context\end{tabular}};

\draw [decorate, decoration={brace,amplitude=5pt,mirror}, xshift=-4pt, yshift=0pt] ($ (auxiliary.south east) + (0, 0.25) $) -- ($ (auxiliary.south west) + (0, 0.25) $) node [black, midway, yshift=20pt, rotate=0] {\normalsize \begin{tabular}{c}auxiliary features for\\current term\end{tabular}};

\end{tikzpicture}}
\vspace*{-1.5\baselineskip}
\caption{The model convolves over the terms in the message. For every term, we represent it using the word representations of its context (learned as part of the model). After creating a representation of the term's context, we concatenate auxiliary features (Table~\ref{tbl:features}). At the output layer, a score is returned as output for every term in the message. The softmax function converts the raw scores to a distribution over the message terms and the \protect\CaptionEndOfRanking{} token. Grayscale intensity in the distribution depicts probability mass.\label{fig:model}}

\end{figure}

\newcommand{\Term}{w}
\newcommand{\ContextWindow}{L}

\begin{table}[t]

\caption{Overview of term representation (learned as part of the model) and auxiliary features.\label{tbl:features}}
\small
\resizebox{\columnwidth}{!}{%
\begin{tabularx}{\linewidth}{lX}
\toprule

\multicolumn{2}{l}{\textbf{Context features (learned representations)}} \\

\texttt{term} & Representation of the term. \\
\texttt{context} & Representations of the context surrounding the term. \\

\midrule

\multicolumn{2}{l}{\textbf{Part-of-Speech features}} \\

\texttt{is\_noun} & POS tagged as a noun \citep{Bendersky2008discoveringkeyconcepts} \\
\texttt{is\_verb} & POS tagged as a verb \\
\texttt{is\_other} & POS tagged as neither a noun or a verb \\

\midrule

\multicolumn{2}{l}{\textbf{Message features}} \\

\texttt{is\_subject} & Term occurrence is part of the subject \citep{Cetintas2012querygenerationpriorart} \\
\texttt{is\_body} & Term occurrence is part of the body \citep{Cetintas2012querygenerationpriorart} \\
\texttt{Abs. TF} & Abs. term freq. within the message \citep{Xue2009querygenerationpatents} \\
\texttt{Rel. TF} & Rel. term freq. within the message \citep{Xue2009querygenerationpatents} \\
\texttt{Rel. pos.} & Rel. position of the term within the message \\
\texttt{is\_oov\_repr} & Term does not have a learned representation \\

\midrule

\multicolumn{2}{l}{\textbf{Collection statistics features}} \\

\texttt{IDF} & Inverse document frequency of the term \citep{Xue2009querygenerationpatents} \\
\texttt{TF-IDF} & \texttt{TF} $\times$ \texttt{IDF} \citep{Xue2009querygenerationpatents} \\

\texttt{Abs. CF} & Abs. collection freq. within the collection \\
\texttt{Rel. CF} & Rel. collection freq. within the collection \\

\texttt{Rel. Entropy} & KL divergence from the unsmoothed collection term distribution to the smoothed ($\lambda = 0.5$) document term distribution \citep{Mahdabi2011buildingqueriespriorart} \\

\texttt{SCQ} & Similarity Collection/Query \citep{Zhao2008sqs} \\
\texttt{ICTF} & Inverse Collection Term Frequency \citep{Kumaran2009reducinglongqueries} \\
\texttt{Pointwise SCS} & Pointwise Simplified Clarity Score \citep{He2004scs} \\

\bottomrule
\end{tabularx}%
}

\end{table}

Our convolutional neural network (\Neural{}) term ranking model is organized as follows; see Fig.~\ref{fig:model} for an overview. Given request message $\RequestMessage{}$, we perform a convolution over the $n$ message terms $\Term{}_1, \ldots, \Term{}_n$. Every term $\Term{}_k$ is characterized by
\begin{inparaenum}[(1)]
	\item the term $\Term{}_k$ itself,
	\item the $2 \cdot \ContextWindow{}$ terms, $\Term{}_{k - \ContextWindow{}}, \ldots, \Term{}_{k - 1}, \Term{}_{k + 1}, \ldots, \Term{}_{k + \ContextWindow{}}$, surrounding term $\Term{}_k$ where $\ContextWindow$ is a context width hyperparameter, and
	\item auxiliary query term quality features (see Table~\ref{tbl:features}).
\end{inparaenum}
\newcommand{\TermRankingFn}{g}
\newcommand{\TermRankingScore}[2]{\Apply{\TermRankingFn}{#1, #2}}
\newcommand{\EndOfRankFn}{h}
\newcommand{\EndOfRankScore}[1]{\Apply{\EndOfRankFn}{#1}}
For every term in the message, the local context features (1st part of Table~\ref{tbl:features}) are looked up in term embedding matrix $\TermEmbeddingMatrix{}$ (learned as part of the model) and the auxiliary features (part 2-4 of Table~\ref{tbl:features}) are computed. For the auxiliary features, we apply min-max feature scaling on the message-level such that they fall between $0$ and $1$. The concatenated embeddings and auxiliary feature vector, are fed to the neural network. At the output layer, the network predicts a term ranking score, $\TermRankingScore{\Term{}_k}{\RequestMessage{}}$, for every term. In addition, a score for the \EndOfRanking{} token, $\EndOfRankScore{\RequestMessage{}}$, is predicted as well. The \EndOfRanking{} score function $\EndOfRankFn{}$ takes the same form as the term score function $\TermRankingFn{}$, but has a separate set of parameters (due to its input features having a different distribution) and takes as input an aggregated vector that represents the whole message. More specifically, the input to the \EndOfRanking{} score function is the average of the term representations and their auxiliary features.
The ranking scores are then transformed into a distribution over message terms and the \EndOfRanking{} token as follows:
\newcommand{\PartitionFn}{\exp{\EndOfRankScore{\RequestMessage{}}} + \sum^{\Length{\RequestMessage{}}}_{l = 1} \exp{\TermRankingScore{\Term{}_l}{\RequestMessage{}}}}
\begin{flalign*}
\small
\CondProb{\Term{}_k}{\RequestMessage{}} = \frac{1}{Z_2} {\exp{\TermRankingScore{\Term{}_k}{\RequestMessage{}}}} \enskip\text{and}\enskip
\CondProb{\EndOfRanking{}}{\RequestMessage{}} = \frac{1}{Z_2} {\exp{\EndOfRankScore{\RequestMessage{}}}}
\end{flalign*}
with
$
\Apply{Z_2}{\RequestMessage} = \PartitionFn{}
$ as a normalization constant that normalizes the raw term scores such that we obtain a distribution over message terms.
\newcommand{\NormalizedGroundTruthProb}[1]{\Apply{Q}{#1}}
\newcommand{\UnnormalizedGroundTruthProb}[1]{\Apply{\hat{Q}}{#1}}
For every query $\CandidateQuery{} \in \CandidateQueries{}$, the ground-truth distribution equals:
\begin{flalign}
\label{eq:rank_groundtruth}
\small
\CondProb[\NormalizedGroundTruthProb]{\Term{}_k}{\CandidateQuery{}} = \alpha \cdot \frac{\Apply{\mathds{1}_{\CandidateQuery{}}}{\Term{}_k}}{\Apply{\FrequencyFn{}}{\Term{}_k, \RequestMessage{}} \cdot \Length{\CandidateQuery{}}}  \quad\text{ and }\quad
\CondProb[\NormalizedGroundTruthProb]{\EndOfRanking{}}{\CandidateQuery{}} = (1 - \alpha)
\end{flalign}
where $\alpha = 0.95$ is a hyperparameter that determines the probability mass assigned to the \EndOfRanking{} token and $\Apply{\mathds{1}_{\CandidateQuery{}}}{\Term{}_k}$ is the indicator function that evaluates to $1$ when term $\Term{}_k$ is part of silver query $\CandidateQuery{}$. The frequency count $\Apply{\FrequencyFn{}}{\Term{}_k, \RequestMessage{}}$ denotes the number of times term $\Term{}_k$ occurs in message $\RequestMessage{}$ and is included such that frequent and infrequent message terms are equally important.
Eq.~\ref{eq:rank_groundtruth} assigns an equal probability to every unique term in message $\RequestMessage{}$ that occurs in silver query $\CandidateQuery{}$. Our cost function consists of two objectives. The first objective aims to make $\CondProb{\;\cdot}{\RequestMessage{}}$ close to $\CondProb[\NormalizedGroundTruthProb]{\;\cdot}{\CandidateQuery{}}$ by minimizing the cross entropy:
\newcommand{\CELossFn}{\LossFn{}_{\text{xent}}}
\newcommand{\CutOffLossFn}{\LossFn{}_{\text{cutoff}}}
\newcommand{\CELoss}[1][\RequestMessage{}]{\Apply{\CELossFn{}}{\Parameters{} \mid #1, \CandidateQuery{}}}
\begin{equation}
\label{eq:ce_loss_fn}
\CELoss{} = - \sum_{\omega \in \Omega} \CondProb[\NormalizedGroundTruthProb]{\omega}{\CandidateQuery{}} \log{\CondProb{\omega}{\RequestMessage{}}}
\end{equation}
where $\Omega = \left(\Term{}_1, \ldots, \Term{}_n, \EndOfRanking{}\right)$ is the sequence of all terms in the message $\RequestMessage{}$ concatenated with the end-of-ranking token. Eq.~\ref{eq:ce_loss_fn} promotes term ranking precision as it causes terms in the silver query to be ranked highly, immediately followed by the end-of-ranking token. The second objective encourages term ranking recall by dictating that the \EndOfRanking{} token should occur at the same rank as the lowest-ranked silver query term:
\newcommand{\CutOffLoss}[1][\RequestMessage{}]{\Apply{\CutOffLossFn{}}{\Parameters{} \mid #1, \CandidateQuery{}}}
\begin{equation}
\label{eq:cutoff_loss_fn}
\CutOffLoss{} = \left(\Apply{\text{min}_{\Term{} \in \CandidateQuery{}}}{\TermRankingScore{\Term{}}{\RequestMessage{}}} - \EndOfRankScore{\RequestMessage{}}\right)^2
\end{equation}
The two objectives (Eq.~\ref{eq:ce_loss_fn}-\ref{eq:cutoff_loss_fn}) are then combined in a batch objective:
\newcommand{\Batch}{B}
\begin{equation}
\label{eq:loss}
\begin{aligned}
\Apply{\LossFn{}}{\Parameters{} \mid \Batch{}} = \frac{1}{\Length{\Batch{}}} \sum_{\left(\Message{}, \CandidateQuery{}\right) \in \Batch{}} \CandidateQueryScore{\CandidateQuery{}} \big( \;\;\;\;\;\;\;\;\;\;\;\;\;\;\;\;\;\;\;\;\;\;\;\;\;\;\;\;\;\;\;\;\;\;\;\;\;\;& \\[-0.75em]
\CELoss[\Message{}]{} + \CutOffLoss[\Message{}]{} \big) + \frac{1}{2 \lambda} \sum_{W \in \Parameters{}_W} \sum_{ij} W^2_{ij} &
\end{aligned}
\end{equation}%
where $\Batch{}$ is a uniformly random sampled batch of message/query pairs, $\Parameters{}_W$ is the set of parameter matrices and $\lambda$ is a weight regularization parameter. Objective~\ref{eq:ce_loss_fn} resembles a list-wise learning to rank method \citep{Cao2007listwise} where a softmax over the top-ranked items is used. Eq.~\ref{eq:loss} is then optimized using gradient descent.


\section{Experimental set-up}
\label{sec:experiment}

\subsection{Research questions}
\label{sec:rq}

\newcommand{\RQ}[2]{%
    \begin{description}[topsep=2pt,leftmargin=0.8cm]%
    \phantomsection\label{section:setup:rq#1}%
    \item[RQ#1] #2%
    \end{description}%
}

\newcommand{\RQRef}[1]{\textbf{\hyperref[section:setup:rq#1]{RQ#1}}}

Having described our proactive email attachment recommendation framework and query formulation model, we now detail the three research question that we seek to answer.

\RQ{1}{\ResearchQuestionOne{}}
What if we consider the different fields (subject and body) in the email message when selecting query terms? To what extent do methods based on selecting the top ranked terms according to term scoring methods (e.g., \TFIDF{}, \RelativeEntropy{}) perform? Can \Neural{}s outperform state-of-the-art learning to rank methods? What can we say about the length of the queries extracted by the different methods?

\RQ{2}{\ResearchQuestionTwo{}}
In the case that \Neural{}s improve retrieval effectiveness over query extraction methods: what can we say about the errors made by \Neural{}s? In particular, in what cases do our deep convolutional neural networks perform better or worse compared to the query term ranking methods under comparison?

\RQ{3}{\ResearchQuestionThree{}}
Are all types of features useful? Can we make any inferences about the email domain or the attachable \entity{} recommendation task?

\subsection{Experimental design}
\label{sec:design}

We operate under the assumption that an incoming message has been identified as a request for content. A query is then formulated from the message using one of the query formulation methods (\S\ref{sec:baselines}). To answer the research questions posed in \S\ref{sec:rq}, we compare \Neural{}s with existing state-of-the-art query term selection methods on enterprise email collections (\RQRef{1}). In addition, we look at the query lengths generated by the formulation methods that perform best. \RQRef{2} is answered by examining the per-instance difference in reciprocal rank (\S\ref{sec:metrics}) and a qualitative analysis where we examine the outlier examples. For \RQRef{3} we perform a feature ablation study where we systematically leave out a feature category (Table~\ref{tbl:features}).

\subsection{Data collections and pre-processing}
\label{sec:collections}

\begin{table}[t!]
\caption{Overview of the enterprise email collections used in this paper: \AvocadoCollection{} (public) and \InternalCollection{} (proprietary).\label{tbl:statistics}}
\resizebox{0.90\columnwidth}{!}{%
\begin{tabular}{lll}%
\toprule%
&\AvocadoCollection&\InternalCollection\\%
\midrule%
\textbf{Messages}&\nprounddigits{0}%
\npdecimalsign{.}%
\npfourdigitnosep%
\numprint{928992.000000}&\nprounddigits{0}%
\npdecimalsign{.}%
\npfourdigitnosep%
\numprint{1047311.000000}\\%
Message length (terms)&\nprounddigits{2}%
\npdecimalsign{.}%
\npfourdigitnosep%
\numprint{112.333598}%
$\, \pm \,$%
\nprounddigits{2}%
\npdecimalsign{.}%
\npfourdigitnosep%
\numprint{244.009781}&\nprounddigits{2}%
\npdecimalsign{.}%
\npfourdigitnosep%
\numprint{74.696523}%
$\, \pm \,$%
\nprounddigits{2}%
\npdecimalsign{.}%
\npfourdigitnosep%
\numprint{551.884079}\\%
Threads&\nprounddigits{0}%
\npdecimalsign{.}%
\npfourdigitnosep%
\numprint{804010.000000}&\nprounddigits{0}%
\npdecimalsign{.}%
\npfourdigitnosep%
\numprint{381448.000000}\\%
Thread lengths&\nprounddigits{2}%
\npdecimalsign{.}%
\npfourdigitnosep%
\numprint{1.194065}%
$\, \pm \,$%
\nprounddigits{2}%
\npdecimalsign{.}%
\npfourdigitnosep%
\numprint{0.695663}&\nprounddigits{2}%
\npdecimalsign{.}%
\npfourdigitnosep%
\numprint{2.745619}%
$\, \pm \,$%
\nprounddigits{2}%
\npdecimalsign{.}%
\npfourdigitnosep%
\numprint{3.652553}\\%
Time period&3 years, 8 months&1 year\\%
\midrule%
\textbf{Attachable entities}&\nprounddigits{0}%
\npdecimalsign{.}%
\npfourdigitnosep%
\numprint{50462.000000}&\nprounddigits{0}%
\npdecimalsign{.}%
\npfourdigitnosep%
\numprint{28725.000000}\\%
Impressions per \entity{}&\nprounddigits{2}%
\npdecimalsign{.}%
\npfourdigitnosep%
\numprint{3.481313}%
$\, \pm \,$%
\nprounddigits{2}%
\npdecimalsign{.}%
\npfourdigitnosep%
\numprint{2.552939}&\nprounddigits{2}%
\npdecimalsign{.}%
\npfourdigitnosep%
\numprint{2.786736}%
$\, \pm \,$%
\nprounddigits{2}%
\npdecimalsign{.}%
\npfourdigitnosep%
\numprint{1.355253}\\%
\midrule%
\textbf{Messages with an \entity{}}&\nprounddigits{0}%
\npdecimalsign{.}%
\npfourdigitnosep%
\numprint{311478.000000}&\nprounddigits{0}%
\npdecimalsign{.}%
\npfourdigitnosep%
\numprint{152649.000000}\\%
no thread history&\nprounddigits{0}%
\npdecimalsign{.}%
\npfourdigitnosep%
\numprint{288099.000000}&\nprounddigits{0}%
\npdecimalsign{.}%
\npfourdigitnosep%
\numprint{69796.000000}\\%
all \entities{} filtered (\S\ref{sec:collections}) &\nprounddigits{0}%
\npdecimalsign{.}%
\npfourdigitnosep%
\numprint{22399.000000}&\nprounddigits{0}%
\npdecimalsign{.}%
\npfourdigitnosep%
\numprint{80717.000000}\\%
\textbf{Request/reply pairs}&\nprounddigits{0}%
\npdecimalsign{.}%
\npfourdigitnosep%
\numprint{980.000000}&\nprounddigits{0}%
\npdecimalsign{.}%
\npfourdigitnosep%
\numprint{2136.000000}\\%
Thread history length of pairs&\nprounddigits{2}%
\npdecimalsign{.}%
\npfourdigitnosep%
\numprint{1.526531}%
$\, \pm \,$%
\nprounddigits{2}%
\npdecimalsign{.}%
\npfourdigitnosep%
\numprint{1.132232}&\nprounddigits{2}%
\npdecimalsign{.}%
\npfourdigitnosep%
\numprint{4.039794}%
$\, \pm \,$%
\nprounddigits{2}%
\npdecimalsign{.}%
\npfourdigitnosep%
\numprint{5.776973}\\%
Relevant \entities{} per pair&\nprounddigits{2}%
\npdecimalsign{.}%
\npfourdigitnosep%
\numprint{1.221429}%
$\, \pm \,$%
\nprounddigits{2}%
\npdecimalsign{.}%
\npfourdigitnosep%
\numprint{0.704903}&\nprounddigits{2}%
\npdecimalsign{.}%
\npfourdigitnosep%
\numprint{1.294007}%
$\, \pm \,$%
\nprounddigits{2}%
\npdecimalsign{.}%
\npfourdigitnosep%
\numprint{1.821553}\\%
\bottomrule%
\end{tabular}}
\end{table}

We answer our research questions (\S\ref{sec:rq}) using two enterprise email collections that each constitute a single tenant (i.e., an organization):
\begin{inparaenum}[(1)]
\item the \AvocadoCollection{} collection \citep{Oard2015avocado} is a public data set that consists of emails taken from \numprint{279} custodians of a defunct information technology company, and
\item the Proprietary Internal Emails (\InternalCollection{}) collection is a proprietary dataset of Microsoft internal emails obtained through an employee participation program.
\end{inparaenum}
We perform cross-validation on the collection level. That is, when testing on one collection, models are trained and hyperparameters are selected on the other collection (i.e., train/validate on \AvocadoCollection{}, test on \InternalCollection{} and vice versa). Models should generalize over multiple tenants (i.e., organizations) as maintaining specialized models is cumbersome. In addition, model effectiveness should remain constant over time to avoid frequent model retraining. Consequently, topical regularities contained within a tenant should not influence our comparison. Furthermore, privacy concerns may dictate that training and test tenants are different. On the training set, we create a temporal 95/5 split for training and validation/model selection. On the test set, all instances are used for testing only. Attachable \entities{} consist of file attachments and URLs; see Table~\ref{tbl:statistics}.

The training and test instances are extracted, for every collection independently, in the following unsupervised manner. File attachments and normalized URLs are extracted from all messages. We remove outlier \entities{} by trimming the bottom and top 5\% of the attachable \entity{} frequency distribution. Infrequent \entities{} are non-retrievable and are removed in accordance to our experimental design (\S\ref{sec:design}). However, in this paper we are interested in measuring the performance on retrieving attachable \entities{} that are in the ``torso'' of the distribution and, consequently, frequent \entities{} (e.g., electronic business cards) are removed as well. Any message that links to an attachable \entity{} (i.e., URL or attachment) and the message preceding it is considered a request/reply instance. In addition, we filter request/reply instances containing attachable \entities{} that
\begin{inparaenum}[(a)]
	\item occurred previously in the same thread, or
	\item contain attachable \entities{} that do not occur in the user's mailbox before the time of the request message (see \S\ref{sec:design}).
\end{inparaenum}
Mailboxes are indexed and searched using Indri \cite{Strohman2005indri,VanGysel2017pyndri}. For message retrieval, we use the Query-Likelihood Model (QLM) with Dirichlet smoothing \citep{Zhai2004smoothing} where the smoothing parameter ($\mu$) is set to the average message length \citep{Balog2006expertfinding,Weerkamp2009contextualinformationemailsearch}. At test time, query formulation methods extract query terms from the request message, queries are executed using the email search engine of the user (i.e., Indri) and attachable \entities{} are ranked according to Eq.~\ref{eq:ranking}. Rankings are truncated such that they only contain the top-1000 messages and top-100 attachable \entities{}. The ground truth consists of binary labels where \entities{} part of the reply message are relevant.

\newcommand{\RQAnswer}[3]{
\begin{description}[topsep=0pt,parsep=0pt,leftmargin=0.8cm]
	\item[\RQRef{#1}] #2
\end{description}
\noindent #3}

\newcommand{\Significant}{$^{*\phantom{*}}$}
\newcommand{\MoreSignificant}{$^{**}$}

\subsection{Methods under comparison}
\label{sec:baselines}

As the attachable \entity{} recommendation task is first introduced in this paper, there exist no methods directly aimed at solving this task. However, as mentioned in the related work section (\S\ref{sec:related_work}), there are two areas (prior art search and verbose query reduction) that focus on extracting queries from texts. Consequently, we use computationally tractable methods (\S\ref{sec:related_work}) from these areas for comparison:
\newcommand{\Subject}{subject}
\newcommand{\Body}{body}
\begin{inparaenum}[(1)]
	\item single features, i.e., term frequency (\TF{}), \TFIDF{}, \logTFIDF{}, relative entropy (\RelativeEntropy{}), used for prior art retrieval \citep{Xue2009transformingpatents,Xue2009querygenerationpatents,Mahdabi2011buildingqueriespriorart,Cetintas2012querygenerationpriorart} where the top-$k$ unique terms are selected from either the \Subject{}, the \Body{} or both. Hyperparameters $1 \leq k \leq 15$ and, in the case of \RelativeEntropy{}, $\lambda = 0.1, \ldots, 0.9$ are optimized on the validation set,
	\item the learning to rank method for query term ranking proposed by \citet{Lee2009querytermsranking} for the verbose query reduction task. To adapt this method for our purposes, we use the domain-specific features listed in Table~\ref{tbl:features} (where the representations are obtained by training a Skip-Gram word2vec model with default parameters on the email collection), only consider single-term groups (as higher order term groups are computationally impractical during inference) and use a more powerful pairwise \LeaveOutRankSVM{} \citep{Joachims2002optimizing} (with default parameters \citep{Sculley2009ranksvm}) instead of a pointwise approach. Feature value min-max normalization is performed on the instance-level. The context window width $\ContextWindow{} = 3, 5, \ldots, 15$ is optimized on the validation set.
	In addition, we consider the following baselines:
	\item all terms (\AllTerms{}) are selected from either the \Subject{}, the \Body{} or both,
	\item random terms, selected from the \Subject{}, the \Body{} or both, where we either select $k$ unique terms randomly (\Multinomial{}) or a random percentage $p$ of terms (\BernouilliProcess{}). Hyperparameters $1 \leq k \leq 15$ and $p = 10\%, 20\%, \ldots, 50\%$ are optimized on the validation set.
	Finally, we consider a pointwise alternative to the \NeuralRankCutoff{} model:
	\item \NeuralLogistic{} with the logistic function at the output layer (instead of the softmax) and terms are selected if their score exceeds a threshold optimized on the validation set F1 score (instead of the \EndOfRanking{} token).
\end{inparaenum}

The \Neural{} models are trained for \numprint{30} iterations using Adam \citep{Kingma2014adam} with $\alpha = 10^{-5}$, $\beta_1 = 0.9$, $\beta_2 = 0.999$ and $\epsilon = 10^{-8}$; we select the iteration with lowest data loss on the validation set. Word embeddings are \numprint{128}-dim., the two hidden layers have \numprint{512} hidden units each, with dropout ($p = 0.50$) and the softplus function. Weights are initialized according to \citep{Glorot2010}. Batch size $\Length{\Batch{}} = 128$ and regularization lambda $\lambda = 0.1$. The context window width $\ContextWindow{} = 3, 5, \ldots, 15$ is optimized on the validation set. For word embeddings (both as part of the \Neural{} and \LeaveOutRankSVM{}), we consider the top-$60\text{k}$ terms. Infrequent terms share a representation for the unknown token.

\subsection{Evaluation measures}
\label{sec:metrics}

To answer \RQRef{1}, we report the Mean Reciprocal Rank (\RecipRank{}), Normalized Discounted Cumulative Gain (\NDCG{}) and Precision@5 (\PrecisionCut{}) measures.\footnote{Computed using \texttt{trec\_eval}: \url{https://github.com/usnistgov/trec_eval}} For \RQRef{2}, we examine the pairwise differences in terms of Reciprocal Rank (RR). In the case of \RQRef{3}, we measure the relative difference in \RecipRank{} when removing a feature category.


\section{Results \& Discussion}
\label{sec:discussion}

\subsection{Overview of experimental results}

\begin{table}[t!]
\setlength{\tabcolsep}{0pt}
\renewcommand{\arraystretch}{1.0}
\caption{Comparison of \RankModel{} with state-of-the-art query formulation methods (\S\ref{sec:baselines}) on the \AvocadoCollection{} and \InternalCollection{} collections. The numbers reported on \AvocadoCollection{} were obtained using models trained/validated on \InternalCollection{} and vice versa (\S\ref{sec:collections}). Significance is determined using a paired two-tailed Student t-test ($^{*}$ $p < 0.10$; \MoreSignificant{} $p < 0.05$) \citep{Smucker2007significance} between \NeuralRankCutoff{} and the second best performing method (in italic).\label{tbl:results}}
\centering
\resizebox{0.90\columnwidth}{!}{\begin{tabular}{l@{\extracolsep{1em}}c@{\extracolsep{0pt}}cc@{\extracolsep{1em}}c@{\extracolsep{0pt}}cc}%
\toprule%
&\multicolumn{3}{c}{\AvocadoCollection}&\multicolumn{3}{c}{\InternalCollection}\\%
&\RecipRank&\NDCG&\PrecisionCut&\RecipRank&\NDCG&\PrecisionCut\\%
\midrule%
\multicolumn{7}{l}{\textbf{Full field, single features and random (subject)}}\\%
\SubjectAllTerms&\phantom{0}%
\nprounddigits{4}%
\npdecimalsign{.}%
\npfourdigitnosep%
\numprint{0.228597}%
\phantom{\MoreSignificant}&\phantom{0}%
\nprounddigits{4}%
\npdecimalsign{.}%
\npfourdigitnosep%
\numprint{0.309700}%
\phantom{\MoreSignificant}&\phantom{0}%
\nprounddigits{4}%
\npdecimalsign{.}%
\npfourdigitnosep%
\numprint{0.068577}%
\phantom{\MoreSignificant}&\phantom{0}%
\nprounddigits{4}%
\npdecimalsign{.}%
\npfourdigitnosep%
\numprint{0.333833}%
\phantom{\MoreSignificant}&\phantom{0}%
\nprounddigits{4}%
\npdecimalsign{.}%
\npfourdigitnosep%
\numprint{0.462139}%
\phantom{\MoreSignificant}&\phantom{0}%
\nprounddigits{4}%
\npdecimalsign{.}%
\npfourdigitnosep%
\numprint{0.108801}%
\phantom{\MoreSignificant}\\%
\SubjectTF&\phantom{0}%
\nprounddigits{4}%
\npdecimalsign{.}%
\npfourdigitnosep%
\numprint{0.228035}%
\phantom{\MoreSignificant}&\phantom{0}%
\nprounddigits{4}%
\npdecimalsign{.}%
\npfourdigitnosep%
\numprint{0.309465}%
\phantom{\MoreSignificant}&\phantom{0}%
\nprounddigits{4}%
\npdecimalsign{.}%
\npfourdigitnosep%
\numprint{0.068577}%
\phantom{\MoreSignificant}&\phantom{0}%
\nprounddigits{4}%
\npdecimalsign{.}%
\npfourdigitnosep%
\numprint{0.331518}%
\phantom{\MoreSignificant}&\phantom{0}%
\nprounddigits{4}%
\npdecimalsign{.}%
\npfourdigitnosep%
\numprint{0.459958}%
\phantom{\MoreSignificant}&\phantom{0}%
\nprounddigits{4}%
\npdecimalsign{.}%
\npfourdigitnosep%
\numprint{0.107865}%
\phantom{\MoreSignificant}\\%
\SubjectTFIDF&\phantom{0}%
\nprounddigits{4}%
\npdecimalsign{.}%
\npfourdigitnosep%
\numprint{0.224964}%
\phantom{\MoreSignificant}&\phantom{0}%
\nprounddigits{4}%
\npdecimalsign{.}%
\npfourdigitnosep%
\numprint{0.307309}%
\phantom{\MoreSignificant}&\phantom{0}%
\nprounddigits{4}%
\npdecimalsign{.}%
\npfourdigitnosep%
\numprint{0.070420}%
\phantom{\MoreSignificant}&\phantom{0}%
\nprounddigits{4}%
\npdecimalsign{.}%
\npfourdigitnosep%
\numprint{0.339043}%
\phantom{\MoreSignificant}&\phantom{0}%
\nprounddigits{4}%
\npdecimalsign{.}%
\npfourdigitnosep%
\numprint{0.466279}%
\phantom{\MoreSignificant}&\phantom{0}%
\nprounddigits{4}%
\npdecimalsign{.}%
\npfourdigitnosep%
\numprint{0.108989}%
\phantom{\MoreSignificant}\\%
\SubjectlogTFIDF&\phantom{0}%
\nprounddigits{4}%
\npdecimalsign{.}%
\npfourdigitnosep%
\numprint{0.228035}%
\phantom{\MoreSignificant}&\phantom{0}%
\nprounddigits{4}%
\npdecimalsign{.}%
\npfourdigitnosep%
\numprint{0.309465}%
\phantom{\MoreSignificant}&\phantom{0}%
\nprounddigits{4}%
\npdecimalsign{.}%
\npfourdigitnosep%
\numprint{0.068577}%
\phantom{\MoreSignificant}&\phantom{0}%
\nprounddigits{4}%
\npdecimalsign{.}%
\npfourdigitnosep%
\numprint{0.331518}%
\phantom{\MoreSignificant}&\phantom{0}%
\nprounddigits{4}%
\npdecimalsign{.}%
\npfourdigitnosep%
\numprint{0.459958}%
\phantom{\MoreSignificant}&\phantom{0}%
\nprounddigits{4}%
\npdecimalsign{.}%
\npfourdigitnosep%
\numprint{0.107865}%
\phantom{\MoreSignificant}\\%
\SubjectRelativeEntropy&\phantom{0}%
\nprounddigits{4}%
\npdecimalsign{.}%
\npfourdigitnosep%
\numprint{0.222337}%
\phantom{\MoreSignificant}&\phantom{0}%
\nprounddigits{4}%
\npdecimalsign{.}%
\npfourdigitnosep%
\numprint{0.303838}%
\phantom{\MoreSignificant}&\phantom{0}%
\nprounddigits{4}%
\npdecimalsign{.}%
\npfourdigitnosep%
\numprint{0.069806}%
\phantom{\MoreSignificant}&\phantom{0}%
\nprounddigits{4}%
\npdecimalsign{.}%
\npfourdigitnosep%
\emph{\numprint{0.339116}}%
\phantom{\MoreSignificant}&\phantom{0}%
\nprounddigits{4}%
\npdecimalsign{.}%
\npfourdigitnosep%
\emph{\numprint{0.466422}}%
\phantom{\MoreSignificant}&\phantom{0}%
\nprounddigits{4}%
\npdecimalsign{.}%
\npfourdigitnosep%
\emph{\numprint{0.109457}}%
\phantom{\MoreSignificant}\\%
\SubjectMultinomial&\phantom{0}%
\nprounddigits{4}%
\npdecimalsign{.}%
\npfourdigitnosep%
\numprint{0.214253}%
\phantom{\MoreSignificant}&\phantom{0}%
\nprounddigits{4}%
\npdecimalsign{.}%
\npfourdigitnosep%
\numprint{0.293195}%
\phantom{\MoreSignificant}&\phantom{0}%
\nprounddigits{4}%
\npdecimalsign{.}%
\npfourdigitnosep%
\numprint{0.064688}%
\phantom{\MoreSignificant}&\phantom{0}%
\nprounddigits{4}%
\npdecimalsign{.}%
\npfourdigitnosep%
\numprint{0.326604}%
\phantom{\MoreSignificant}&\phantom{0}%
\nprounddigits{4}%
\npdecimalsign{.}%
\npfourdigitnosep%
\numprint{0.455315}%
\phantom{\MoreSignificant}&\phantom{0}%
\nprounddigits{4}%
\npdecimalsign{.}%
\npfourdigitnosep%
\numprint{0.106273}%
\phantom{\MoreSignificant}\\%
\SubjectBernouilliProcess&\phantom{0}%
\nprounddigits{4}%
\npdecimalsign{.}%
\npfourdigitnosep%
\numprint{0.148068}%
\phantom{\MoreSignificant}&\phantom{0}%
\nprounddigits{4}%
\npdecimalsign{.}%
\npfourdigitnosep%
\numprint{0.210405}%
\phantom{\MoreSignificant}&\phantom{0}%
\nprounddigits{4}%
\npdecimalsign{.}%
\npfourdigitnosep%
\numprint{0.046673}%
\phantom{\MoreSignificant}&\phantom{0}%
\nprounddigits{4}%
\npdecimalsign{.}%
\npfourdigitnosep%
\numprint{0.274876}%
\phantom{\MoreSignificant}&\phantom{0}%
\nprounddigits{4}%
\npdecimalsign{.}%
\npfourdigitnosep%
\numprint{0.401259}%
\phantom{\MoreSignificant}&\phantom{0}%
\nprounddigits{4}%
\npdecimalsign{.}%
\npfourdigitnosep%
\numprint{0.088858}%
\phantom{\MoreSignificant}\\%
\midrule%
\multicolumn{7}{l}{\textbf{Full field, single features and random (body)}}\\%
\BodyAllTerms&\phantom{0}%
\nprounddigits{4}%
\npdecimalsign{.}%
\npfourdigitnosep%
\numprint{0.124781}%
\phantom{\MoreSignificant}&\phantom{0}%
\nprounddigits{4}%
\npdecimalsign{.}%
\npfourdigitnosep%
\numprint{0.193012}%
\phantom{\MoreSignificant}&\phantom{0}%
\nprounddigits{4}%
\npdecimalsign{.}%
\npfourdigitnosep%
\numprint{0.037666}%
\phantom{\MoreSignificant}&\phantom{0}%
\nprounddigits{4}%
\npdecimalsign{.}%
\npfourdigitnosep%
\numprint{0.211488}%
\phantom{\MoreSignificant}&\phantom{0}%
\nprounddigits{4}%
\npdecimalsign{.}%
\npfourdigitnosep%
\numprint{0.337609}%
\phantom{\MoreSignificant}&\phantom{0}%
\nprounddigits{4}%
\npdecimalsign{.}%
\npfourdigitnosep%
\numprint{0.067228}%
\phantom{\MoreSignificant}\\%
\BodyTF&\phantom{0}%
\nprounddigits{4}%
\npdecimalsign{.}%
\npfourdigitnosep%
\numprint{0.102476}%
\phantom{\MoreSignificant}&\phantom{0}%
\nprounddigits{4}%
\npdecimalsign{.}%
\npfourdigitnosep%
\numprint{0.171940}%
\phantom{\MoreSignificant}&\phantom{0}%
\nprounddigits{4}%
\npdecimalsign{.}%
\npfourdigitnosep%
\numprint{0.030911}%
\phantom{\MoreSignificant}&\phantom{0}%
\nprounddigits{4}%
\npdecimalsign{.}%
\npfourdigitnosep%
\numprint{0.209354}%
\phantom{\MoreSignificant}&\phantom{0}%
\nprounddigits{4}%
\npdecimalsign{.}%
\npfourdigitnosep%
\numprint{0.335756}%
\phantom{\MoreSignificant}&\phantom{0}%
\nprounddigits{4}%
\npdecimalsign{.}%
\npfourdigitnosep%
\numprint{0.066011}%
\phantom{\MoreSignificant}\\%
\BodyTFIDF&\phantom{0}%
\nprounddigits{4}%
\npdecimalsign{.}%
\npfourdigitnosep%
\numprint{0.150676}%
\phantom{\MoreSignificant}&\phantom{0}%
\nprounddigits{4}%
\npdecimalsign{.}%
\npfourdigitnosep%
\numprint{0.221342}%
\phantom{\MoreSignificant}&\phantom{0}%
\nprounddigits{4}%
\npdecimalsign{.}%
\npfourdigitnosep%
\numprint{0.045855}%
\phantom{\MoreSignificant}&\phantom{0}%
\nprounddigits{4}%
\npdecimalsign{.}%
\npfourdigitnosep%
\numprint{0.223661}%
\phantom{\MoreSignificant}&\phantom{0}%
\nprounddigits{4}%
\npdecimalsign{.}%
\npfourdigitnosep%
\numprint{0.348055}%
\phantom{\MoreSignificant}&\phantom{0}%
\nprounddigits{4}%
\npdecimalsign{.}%
\npfourdigitnosep%
\numprint{0.072191}%
\phantom{\MoreSignificant}\\%
\BodylogTFIDF&\phantom{0}%
\nprounddigits{4}%
\npdecimalsign{.}%
\npfourdigitnosep%
\numprint{0.110945}%
\phantom{\MoreSignificant}&\phantom{0}%
\nprounddigits{4}%
\npdecimalsign{.}%
\npfourdigitnosep%
\numprint{0.175549}%
\phantom{\MoreSignificant}&\phantom{0}%
\nprounddigits{4}%
\npdecimalsign{.}%
\npfourdigitnosep%
\numprint{0.031116}%
\phantom{\MoreSignificant}&\phantom{0}%
\nprounddigits{4}%
\npdecimalsign{.}%
\npfourdigitnosep%
\numprint{0.191359}%
\phantom{\MoreSignificant}&\phantom{0}%
\nprounddigits{4}%
\npdecimalsign{.}%
\npfourdigitnosep%
\numprint{0.317975}%
\phantom{\MoreSignificant}&\phantom{0}%
\nprounddigits{4}%
\npdecimalsign{.}%
\npfourdigitnosep%
\numprint{0.062734}%
\phantom{\MoreSignificant}\\%
\BodyRelativeEntropy&\phantom{0}%
\nprounddigits{4}%
\npdecimalsign{.}%
\npfourdigitnosep%
\numprint{0.144132}%
\phantom{\MoreSignificant}&\phantom{0}%
\nprounddigits{4}%
\npdecimalsign{.}%
\npfourdigitnosep%
\numprint{0.212777}%
\phantom{\MoreSignificant}&\phantom{0}%
\nprounddigits{4}%
\npdecimalsign{.}%
\npfourdigitnosep%
\numprint{0.042375}%
\phantom{\MoreSignificant}&\phantom{0}%
\nprounddigits{4}%
\npdecimalsign{.}%
\npfourdigitnosep%
\numprint{0.219798}%
\phantom{\MoreSignificant}&\phantom{0}%
\nprounddigits{4}%
\npdecimalsign{.}%
\npfourdigitnosep%
\numprint{0.343047}%
\phantom{\MoreSignificant}&\phantom{0}%
\nprounddigits{4}%
\npdecimalsign{.}%
\npfourdigitnosep%
\numprint{0.069944}%
\phantom{\MoreSignificant}\\%
\BodyMultinomial&\phantom{0}%
\nprounddigits{4}%
\npdecimalsign{.}%
\npfourdigitnosep%
\numprint{0.078542}%
\phantom{\MoreSignificant}&\phantom{0}%
\nprounddigits{4}%
\npdecimalsign{.}%
\npfourdigitnosep%
\numprint{0.139401}%
\phantom{\MoreSignificant}&\phantom{0}%
\nprounddigits{4}%
\npdecimalsign{.}%
\npfourdigitnosep%
\numprint{0.022927}%
\phantom{\MoreSignificant}&\phantom{0}%
\nprounddigits{4}%
\npdecimalsign{.}%
\npfourdigitnosep%
\numprint{0.178137}%
\phantom{\MoreSignificant}&\phantom{0}%
\nprounddigits{4}%
\npdecimalsign{.}%
\npfourdigitnosep%
\numprint{0.307805}%
\phantom{\MoreSignificant}&\phantom{0}%
\nprounddigits{4}%
\npdecimalsign{.}%
\npfourdigitnosep%
\numprint{0.056835}%
\phantom{\MoreSignificant}\\%
\BodyBernouilliProcess&\phantom{0}%
\nprounddigits{4}%
\npdecimalsign{.}%
\npfourdigitnosep%
\numprint{0.102960}%
\phantom{\MoreSignificant}&\phantom{0}%
\nprounddigits{4}%
\npdecimalsign{.}%
\npfourdigitnosep%
\numprint{0.164649}%
\phantom{\MoreSignificant}&\phantom{0}%
\nprounddigits{4}%
\npdecimalsign{.}%
\npfourdigitnosep%
\numprint{0.032549}%
\phantom{\MoreSignificant}&\phantom{0}%
\nprounddigits{4}%
\npdecimalsign{.}%
\npfourdigitnosep%
\numprint{0.188688}%
\phantom{\MoreSignificant}&\phantom{0}%
\nprounddigits{4}%
\npdecimalsign{.}%
\npfourdigitnosep%
\numprint{0.312826}%
\phantom{\MoreSignificant}&\phantom{0}%
\nprounddigits{4}%
\npdecimalsign{.}%
\npfourdigitnosep%
\numprint{0.060581}%
\phantom{\MoreSignificant}\\%
\midrule%
\multicolumn{7}{l}{\textbf{Full field, single features and random (subject + body)}}\\%
\SubjectBodyAllTerms&\phantom{0}%
\nprounddigits{4}%
\npdecimalsign{.}%
\npfourdigitnosep%
\numprint{0.199467}%
\phantom{\MoreSignificant}&\phantom{0}%
\nprounddigits{4}%
\npdecimalsign{.}%
\npfourdigitnosep%
\numprint{0.278460}%
\phantom{\MoreSignificant}&\phantom{0}%
\nprounddigits{4}%
\npdecimalsign{.}%
\npfourdigitnosep%
\numprint{0.061208}%
\phantom{\MoreSignificant}&\phantom{0}%
\nprounddigits{4}%
\npdecimalsign{.}%
\npfourdigitnosep%
\numprint{0.308749}%
\phantom{\MoreSignificant}&\phantom{0}%
\nprounddigits{4}%
\npdecimalsign{.}%
\npfourdigitnosep%
\numprint{0.440585}%
\phantom{\MoreSignificant}&\phantom{0}%
\nprounddigits{4}%
\npdecimalsign{.}%
\npfourdigitnosep%
\numprint{0.097191}%
\phantom{\MoreSignificant}\\%
\SubjectBodyTF&\phantom{0}%
\nprounddigits{4}%
\npdecimalsign{.}%
\npfourdigitnosep%
\numprint{0.178331}%
\phantom{\MoreSignificant}&\phantom{0}%
\nprounddigits{4}%
\npdecimalsign{.}%
\npfourdigitnosep%
\numprint{0.265284}%
\phantom{\MoreSignificant}&\phantom{0}%
\nprounddigits{4}%
\npdecimalsign{.}%
\npfourdigitnosep%
\numprint{0.055067}%
\phantom{\MoreSignificant}&\phantom{0}%
\nprounddigits{4}%
\npdecimalsign{.}%
\npfourdigitnosep%
\numprint{0.300549}%
\phantom{\MoreSignificant}&\phantom{0}%
\nprounddigits{4}%
\npdecimalsign{.}%
\npfourdigitnosep%
\numprint{0.433430}%
\phantom{\MoreSignificant}&\phantom{0}%
\nprounddigits{4}%
\npdecimalsign{.}%
\npfourdigitnosep%
\numprint{0.095318}%
\phantom{\MoreSignificant}\\%
\SubjectBodyTFIDF&\phantom{0}%
\nprounddigits{4}%
\npdecimalsign{.}%
\npfourdigitnosep%
\numprint{0.209690}%
\phantom{\MoreSignificant}&\phantom{0}%
\nprounddigits{4}%
\npdecimalsign{.}%
\npfourdigitnosep%
\numprint{0.293303}%
\phantom{\MoreSignificant}&\phantom{0}%
\nprounddigits{4}%
\npdecimalsign{.}%
\npfourdigitnosep%
\numprint{0.064893}%
\phantom{\MoreSignificant}&\phantom{0}%
\nprounddigits{4}%
\npdecimalsign{.}%
\npfourdigitnosep%
\numprint{0.310026}%
\phantom{\MoreSignificant}&\phantom{0}%
\nprounddigits{4}%
\npdecimalsign{.}%
\npfourdigitnosep%
\numprint{0.439683}%
\phantom{\MoreSignificant}&\phantom{0}%
\nprounddigits{4}%
\npdecimalsign{.}%
\npfourdigitnosep%
\numprint{0.099064}%
\phantom{\MoreSignificant}\\%
\SubjectBodylogTFIDF&\phantom{0}%
\nprounddigits{4}%
\npdecimalsign{.}%
\npfourdigitnosep%
\numprint{0.185820}%
\phantom{\MoreSignificant}&\phantom{0}%
\nprounddigits{4}%
\npdecimalsign{.}%
\npfourdigitnosep%
\numprint{0.272588}%
\phantom{\MoreSignificant}&\phantom{0}%
\nprounddigits{4}%
\npdecimalsign{.}%
\npfourdigitnosep%
\numprint{0.059161}%
\phantom{\MoreSignificant}&\phantom{0}%
\nprounddigits{4}%
\npdecimalsign{.}%
\npfourdigitnosep%
\numprint{0.274714}%
\phantom{\MoreSignificant}&\phantom{0}%
\nprounddigits{4}%
\npdecimalsign{.}%
\npfourdigitnosep%
\numprint{0.409768}%
\phantom{\MoreSignificant}&\phantom{0}%
\nprounddigits{4}%
\npdecimalsign{.}%
\npfourdigitnosep%
\numprint{0.087079}%
\phantom{\MoreSignificant}\\%
\SubjectBodyRelativeEntropy&\phantom{0}%
\nprounddigits{4}%
\npdecimalsign{.}%
\npfourdigitnosep%
\numprint{0.213828}%
\phantom{\MoreSignificant}&\phantom{0}%
\nprounddigits{4}%
\npdecimalsign{.}%
\npfourdigitnosep%
\numprint{0.297972}%
\phantom{\MoreSignificant}&\phantom{0}%
\nprounddigits{4}%
\npdecimalsign{.}%
\npfourdigitnosep%
\numprint{0.064893}%
\phantom{\MoreSignificant}&\phantom{0}%
\nprounddigits{4}%
\npdecimalsign{.}%
\npfourdigitnosep%
\numprint{0.320013}%
\phantom{\MoreSignificant}&\phantom{0}%
\nprounddigits{4}%
\npdecimalsign{.}%
\npfourdigitnosep%
\numprint{0.448947}%
\phantom{\MoreSignificant}&\phantom{0}%
\nprounddigits{4}%
\npdecimalsign{.}%
\npfourdigitnosep%
\numprint{0.102341}%
\phantom{\MoreSignificant}\\%
\SubjectBodyMultinomial&\phantom{0}%
\nprounddigits{4}%
\npdecimalsign{.}%
\npfourdigitnosep%
\numprint{0.140385}%
\phantom{\MoreSignificant}&\phantom{0}%
\nprounddigits{4}%
\npdecimalsign{.}%
\npfourdigitnosep%
\numprint{0.214833}%
\phantom{\MoreSignificant}&\phantom{0}%
\nprounddigits{4}%
\npdecimalsign{.}%
\npfourdigitnosep%
\numprint{0.043603}%
\phantom{\MoreSignificant}&\phantom{0}%
\nprounddigits{4}%
\npdecimalsign{.}%
\npfourdigitnosep%
\numprint{0.272083}%
\phantom{\MoreSignificant}&\phantom{0}%
\nprounddigits{4}%
\npdecimalsign{.}%
\npfourdigitnosep%
\numprint{0.407603}%
\phantom{\MoreSignificant}&\phantom{0}%
\nprounddigits{4}%
\npdecimalsign{.}%
\npfourdigitnosep%
\numprint{0.088577}%
\phantom{\MoreSignificant}\\%
\SubjectBodyBernouilliProcess&\phantom{0}%
\nprounddigits{4}%
\npdecimalsign{.}%
\npfourdigitnosep%
\numprint{0.175265}%
\phantom{\MoreSignificant}&\phantom{0}%
\nprounddigits{4}%
\npdecimalsign{.}%
\npfourdigitnosep%
\numprint{0.251404}%
\phantom{\MoreSignificant}&\phantom{0}%
\nprounddigits{4}%
\npdecimalsign{.}%
\npfourdigitnosep%
\numprint{0.051996}%
\phantom{\MoreSignificant}&\phantom{0}%
\nprounddigits{4}%
\npdecimalsign{.}%
\npfourdigitnosep%
\numprint{0.259179}%
\phantom{\MoreSignificant}&\phantom{0}%
\nprounddigits{4}%
\npdecimalsign{.}%
\npfourdigitnosep%
\numprint{0.394055}%
\phantom{\MoreSignificant}&\phantom{0}%
\nprounddigits{4}%
\npdecimalsign{.}%
\npfourdigitnosep%
\numprint{0.082210}%
\phantom{\MoreSignificant}\\%
\midrule%
\multicolumn{7}{l}{\textbf{Learning-to-rank methods (subject + body)}}\\%
\LeaveOutRankSVM&\phantom{0}%
\nprounddigits{4}%
\npdecimalsign{.}%
\npfourdigitnosep%
\numprint{0.165009}%
\phantom{\MoreSignificant}&\phantom{0}%
\nprounddigits{4}%
\npdecimalsign{.}%
\npfourdigitnosep%
\numprint{0.242479}%
\phantom{\MoreSignificant}&\phantom{0}%
\nprounddigits{4}%
\npdecimalsign{.}%
\npfourdigitnosep%
\numprint{0.049744}%
\phantom{\MoreSignificant}&\phantom{0}%
\nprounddigits{4}%
\npdecimalsign{.}%
\npfourdigitnosep%
\numprint{0.307872}%
\phantom{\MoreSignificant}&\phantom{0}%
\nprounddigits{4}%
\npdecimalsign{.}%
\npfourdigitnosep%
\numprint{0.439175}%
\phantom{\MoreSignificant}&\phantom{0}%
\nprounddigits{4}%
\npdecimalsign{.}%
\npfourdigitnosep%
\numprint{0.098034}%
\phantom{\MoreSignificant}\\%
\NeuralLogistic&\phantom{0}%
\nprounddigits{4}%
\npdecimalsign{.}%
\npfourdigitnosep%
\emph{\numprint{0.231941}}%
\phantom{\MoreSignificant}&\phantom{0}%
\nprounddigits{4}%
\npdecimalsign{.}%
\npfourdigitnosep%
\emph{\numprint{0.312931}}%
\phantom{\MoreSignificant}&\phantom{0}%
\nprounddigits{4}%
\npdecimalsign{.}%
\npfourdigitnosep%
\emph{\numprint{0.070829}}%
\phantom{\MoreSignificant}&\phantom{0}%
\nprounddigits{4}%
\npdecimalsign{.}%
\npfourdigitnosep%
\numprint{0.334708}%
\phantom{\MoreSignificant}&\phantom{0}%
\nprounddigits{4}%
\npdecimalsign{.}%
\npfourdigitnosep%
\numprint{0.462987}%
\phantom{\MoreSignificant}&\phantom{0}%
\nprounddigits{4}%
\npdecimalsign{.}%
\npfourdigitnosep%
\numprint{0.108708}%
\phantom{\MoreSignificant}\\%
\NeuralRankCutoff&\phantom{0}%
\nprounddigits{4}%
\npdecimalsign{.}%
\npfourdigitnosep%
\textbf{\numprint{0.245514}}%
\Significant&\phantom{0}%
\nprounddigits{4}%
\npdecimalsign{.}%
\npfourdigitnosep%
\textbf{\numprint{0.331334}}%
\MoreSignificant&\phantom{0}%
\nprounddigits{4}%
\npdecimalsign{.}%
\npfourdigitnosep%
\textbf{\numprint{0.076970}}%
\MoreSignificant&\phantom{0}%
\nprounddigits{4}%
\npdecimalsign{.}%
\npfourdigitnosep%
\textbf{\numprint{0.349190}}%
\MoreSignificant&\phantom{0}%
\nprounddigits{4}%
\npdecimalsign{.}%
\npfourdigitnosep%
\textbf{\numprint{0.474388}}%
\MoreSignificant&\phantom{0}%
\nprounddigits{4}%
\npdecimalsign{.}%
\npfourdigitnosep%
\textbf{\numprint{0.112266}}%
\phantom{\MoreSignificant}\\%
\bottomrule%
\end{tabular}}
\end{table}
\begin{figure*}[t!]
\centering
\begin{subfigure}[b]{\columnwidth}
    \centering
    \includegraphics[width=0.75\textwidth]{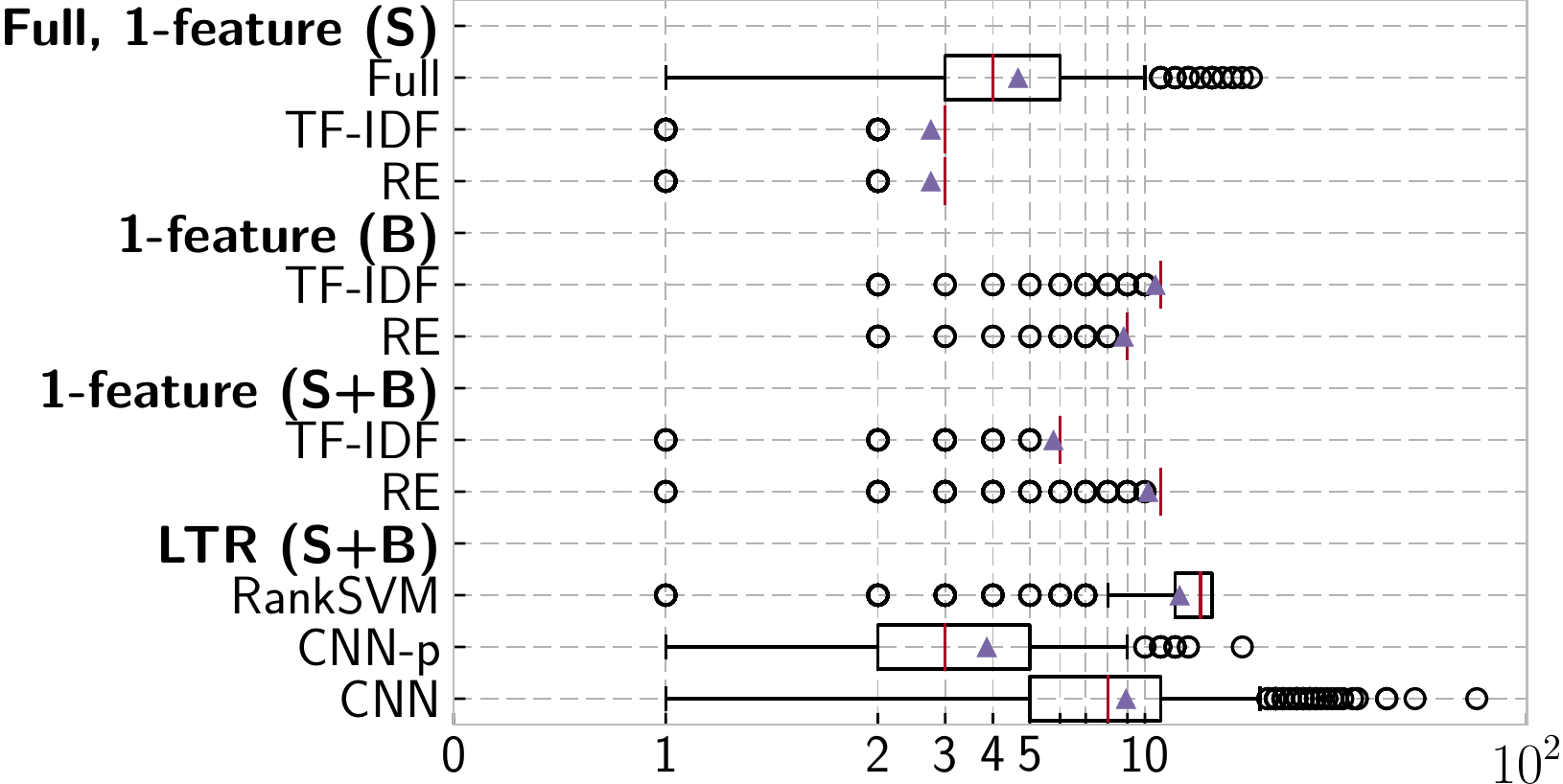}
    \caption{\AvocadoCollection{}}
    \label{fig:query_boxplot:avocado}
\end{subfigure}
~
\begin{subfigure}[b]{\columnwidth}
    \centering
    \includegraphics[width=0.75\textwidth]{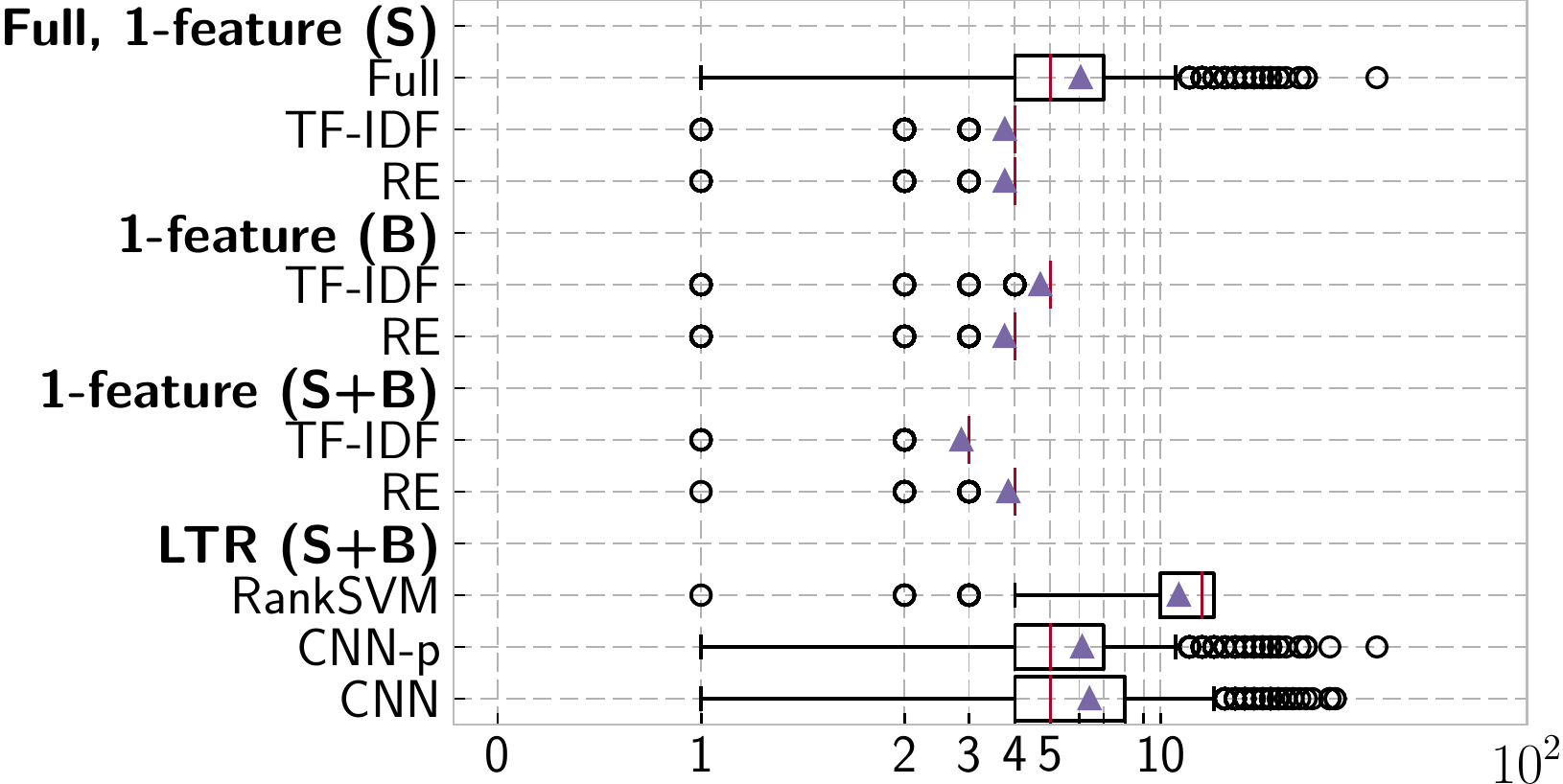}
    \caption{\InternalCollection{}}
    \label{fig:query_boxplot:internal}
\end{subfigure}
\vspace*{-\baselineskip}
\caption{Query length distribution according to the most prominent methods (\textbf{S} and \textbf{B} denote subject and body fields, resp.). \TFIDF{}, \RelativeEntropy{} and \LeaveOutRankSVM{} select a fixed number of terms for all queries, whereas \Neural{}s select a variable number of terms.\label{fig:query_boxplot}}
\end{figure*}

\RQAnswer{1}{Table~\ref{tbl:results} shows the recommendation results of attachable \entities{} in enterprise email collections (\S\ref{sec:collections}).}{%
We see that \RankModel{} outperforms all other query formulation methods on both enterprise email collections. Significance is achieved (\RecipRank{}) between \RankModel{} and the second-best performing methods: \NeuralLogistic{} and \SubjectRelativeEntropy{} (subject), respectively, on the \AvocadoCollection{} and \InternalCollection{}. The methods that select terms only from the email subject perform strongly on both collections. Within the set of subject methods (1st part of Table~\ref{tbl:results}), we also observe that there is little difference between the methods. In fact, for \AvocadoCollection{}, simply taking the subject as query performs better than any of the remaining subject-based methods. Subjects convey relevance and context \citep{Weil2004overcomingoverload} and can compactly describe the topic of a content request. However, in order to generate better queries, we need to extract additional terms from the email body as email subjects tend to be short.

The same methods that we used to extract terms from the subject perform poorly when only presented with the body of the email (2nd part of Table~\ref{tbl:results}). When allowing the methods to select terms from the full email (subject and body), we see a small increase in retrieval performance (3rd part of Table~\ref{tbl:results}) compared to body-only terms. However, none of the methods operating on the full email message manage to outperform the subject by itself.

Our learning to rank (LTR) query term methods (last part of Table~\ref{tbl:results}) outperform the subject-based methods (ignoring \LeaveOutRankSVM{}). This comes as little surprise, as the presence of a term in the subject is incorporated as a feature in our models (Table~\ref{tbl:features}). The reason why \LeaveOutRankSVM{}, originally introduced for reducing long search queries, performs poorly is due to the fact that its training procedure fails to deal with the long length of emails. That is, \LeaveOutRankSVM{} is trained by measuring the decrease in retrieval effectiveness that occurs from leaving one term out of the full email message (i.e., top-down). This is an error-prone way to score query terms as emails tend to be relatively long (Table~\ref{tbl:statistics}).
Conversely, our approach to generate silver query training data (\S\ref{sec:model:subquery-generation}) considers groups of query terms and the reference query is constructed in a bottom-up fashion.

Fig.~\ref{fig:query_boxplot} shows the distribution of generated query lengths for the most prominent methods (\TFIDF{}, \RelativeEntropy{}, \LeaveOutRankSVM{} and the neural networks). On both collections, subject-oriented methods (\TFIDF{}, \RelativeEntropy{}) seem to extract queries of nearly all the same length. When considering the full subject field, we see that its length varies greatly with outliers of up to 70 query terms. Methods that extract terms from the email body or the full email generate slightly longer queries than the \TFIDF{} and \RelativeEntropy{} subject methods. The methods that learn to rank query terms generate the longest queries. While \LeaveOutRankSVM{} selects query terms according to a global rank cut-off, the \Neural{}s select a variable number of terms for every request message. Consequently, it comes as no surprise that we observe high variance within the \Neural{}-generated query lengths. In addition, we observe that \Neural{}s have a similar query length distribution as the subject queries. This is due to the fact that the \Neural{}s actually expand the subject terms, as we will see in the next section.}

\subsection{Analysis of differences}

\begin{figure}[t]
\centering
\begin{subfigure}[b]{\columnwidth}
    \centering
    \includegraphics[width=0.85\textwidth]{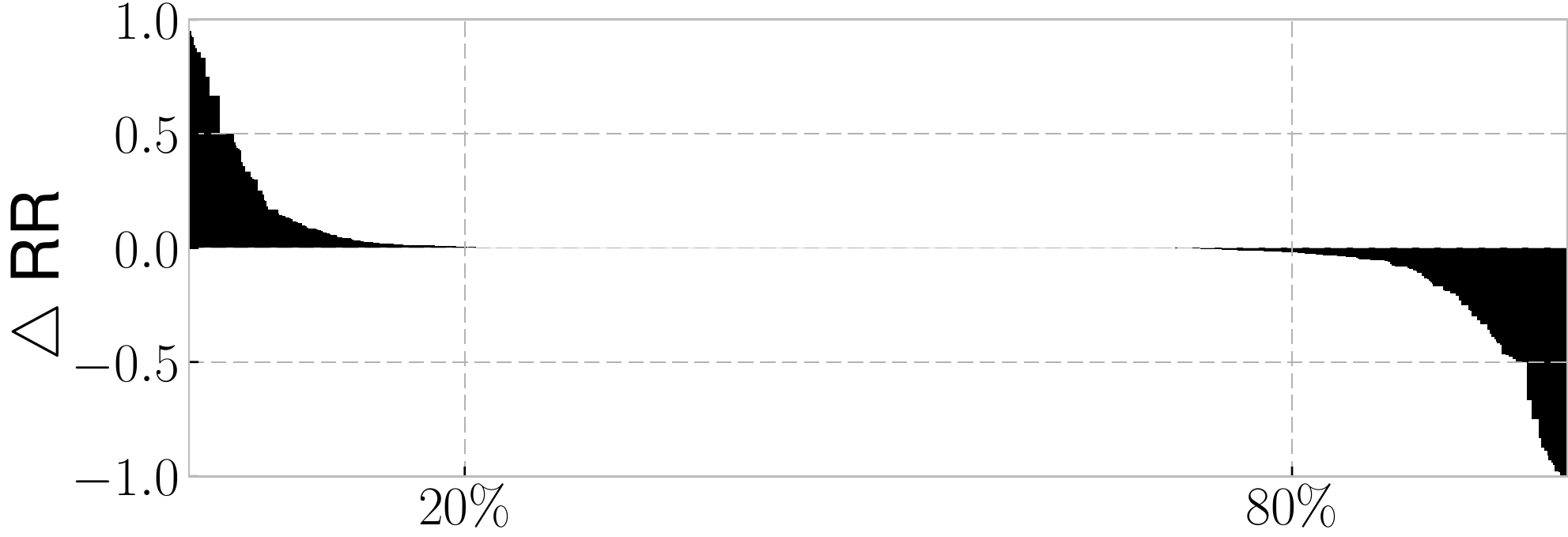}
\end{subfigure}%
\vspace*{-\baselineskip}
\caption{Per-instance differences in Reciprocal Rank (RR) between the email subject query and the \NeuralRankCutoff{} query on \AvocadoCollection{}. The plot for \InternalCollection{} is qualitatively similar. Positive bars (left) indicate instances where the subject performs better than \NeuralRankCutoff{} and vice versa for negative (right).\label{fig:delta}}

\end{figure}

\RQAnswer{2}{Fig.~\ref{fig:delta} shows the per-instance differences between \NeuralRankCutoff{} and the full email subject as query.}{%
For about \numprint{60}\% of request messages both query generation methods (full \Subject{} and \Neural{}) generate queries that perform equally good. In fact, \RecipRank{} on this subset of instances is \numprint{9}\% (\AvocadoCollection{}) and \numprint{17}\% (\InternalCollection{}) better than the best performance over the full test set (Table~\ref{tbl:results}). Do both methods generate identical queries on this subset? The average Jaccard similarity between the queries extracted by both methods is $0.55$ (\AvocadoCollection{}) and $0.62$ (\InternalCollection{}). This indicates that, while query terms extracted by either method overlap, there is a difference in query terms that does not impact retrieval. Upon examining the differences we find that, for the subject, these terms are stopwords and email subject abbreviations (e.g., RE indicating a reply). In the case of \Neural{}, the difference comes from email body terms that further clarify the request. We find that the \Neural{} builds upon the subject query (excluding stopwords) using terms of the body. However, for the \numprint{60}\% of request instances with no difference in performance (Fig.~\ref{fig:delta}), the subject suffices to describe the request.

What can we say about the remaining \numprint{40}\% of queries where there is an observable difference? A closer look at Fig.~\ref{fig:delta} shows that there are extreme peaks at both sides of the graph and that neither method fully dominates the other. Upon examining the outliers, we find the following trends:
\begin{inparaenum}[(1)]
	\item When the subject is indescriptive of the email content (e.g., the subject is \emph{``important issue''}) then the \Neural{} can extract better terms from the email body.
	\item Topic drift within conversations negatively impacts the retrieval effectiveness of the subject as a query. However, long threads do not necessarily exhibit topic drift as in some cases the subject remains a topical representation of the conversation.
	\item Mentions of named entities constitute effective query terms. For example, the name of a person who is responsible for an attachable item tends to improve retrieval effectiveness over using the subject as a query,
	\item Long queries generated by the \Neural{} can disambiguate a request and perform much better than the subject query.
	\item In most cases where the subject query outperforms the \Neural{}, this is due to the fact that the \Neural{} model extracts too many noisy terms and creates query drift.
\end{inparaenum}}

\subsection{Feature importance}

\begin{figure}[t]

\includegraphics[width=0.85\columnwidth]{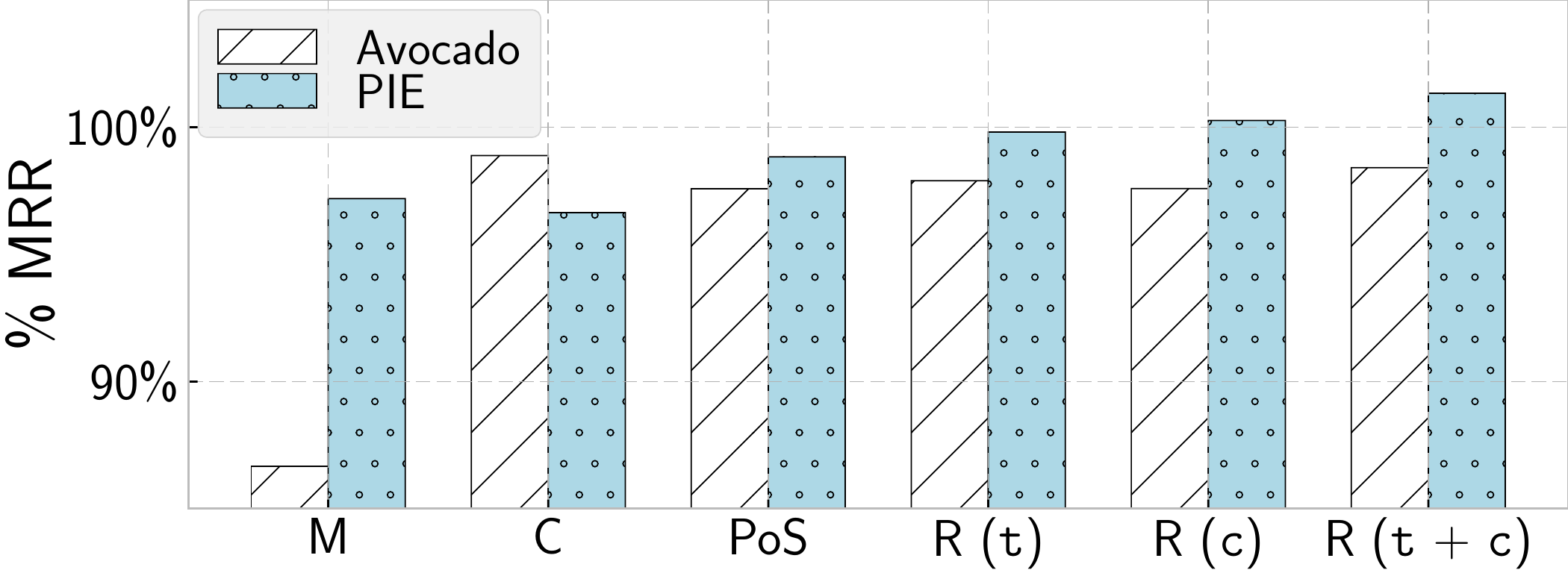}
\vspace*{-\baselineskip}
\caption{Feature ablation study for the \NeuralRankCutoff{} model on the \AvocadoCollection{} (stripes) and \InternalCollection{} (dots) benchmarks. One of the following feature categories (Table~\ref{tbl:features}) is systematically left out: Message (M), Collection statistics (C), Part-of-Speech (PoS), \texttt{term} (\texttt{t}), \texttt{context} (\texttt{c}) or all representations (\texttt{t} + \texttt{c}). \label{fig:feature_ablation}}

\end{figure}

\RQAnswer{3}{Fig.~\ref{fig:feature_ablation} depicts a feature ablation study where we systematically leave out one of the feature categories.}{%
We observe that both local (message and part-of-speech) and global (collection) features are of importance. When comparing the behavior of the enterprise email collections (\S\ref{sec:collections}), we see that the message (M) features have a significant ($p < 0.10$) impact on both collections. The collection statistics (C) yield a significant ($p<0.10$) difference in the case of \InternalCollection{}; no significant differences were observed in the remaining cases. In addition, while \AvocadoCollection{} benefits from the learned representations, the inclusion of the representations of the context slightly decreases performance on \InternalCollection{}. This can be explained by our evaluation setup (\S\ref{sec:design}) where models evaluated on \InternalCollection{} are trained using \AvocadoCollection{} and vice versa. Therefore, it is likely that the model learns certain patterns present from the data-scarce \AvocadoCollection{} collection (Table~\ref{tbl:statistics}) that causes false positive terms to be selected for \InternalCollection{}.}


\section{Conclusions}
\label{sec:conclusion}

We introduced a novel proactive retrieval task for recommending email attachments that involves formulating a query from an email request message. An evaluation framework was proposed that extracts labeled request/attachment instances from an email corpus containing request/reply pairs automatically. Candidate queries, which we refer to as \emph{silver queries}, are synthesized for request/attachment instances and a deep convolutional neural network (\Neural{}) is trained using the silver queries that learns to extract query terms from request messages.
We find that our framework extracts instances that are usable for training and testing. Our \Neural{}, which we train using silver queries, significantly outperforms existing methods for extracting query terms from verbose texts. Terms occurring in the subject of the email are representative of the request and formulating a query using the subject is a strong baseline. A study of the per-instance \RecipRank{} differences show that the \Neural{} and subject query perform quite differently for $40\%$ of instances. A qualitative analysis suggests that our \Neural{} outperforms the subject query in cases where the subject is indescriptive. In addition, mentions of named entities constitute good query terms and lengthy queries disambiguate the request. In cases when the subject query outperforms the \Neural{}, it is due to noisy terms being selected from the email body. A feature ablation study shows that both local (i.e., message) and global (i.e., collection) features are important.

Our work has the following limitations.
\begin{inparaenum}[(1)]
\item In this paper we only consider terms occurring in the request message as candidates. While this prevents the term candidate set from becoming too large, it does limit the ability for methods to formulate expressive queries in the case where request messages are concise.
\item The retrieval model used in this paper, a language model with Dirichlet smoothing, is ubiquitous in retrieval systems. However, smoothing allows the search engine to deal with verbose queries \citep{Zhai2004smoothing} that contain terms absent from the messages. Subjects often contain superfluous terms (e.g., email clients prepend \emph{FW} to the subjects of forwarded messages). Consequently, our findings may change when considering other retrieval model classes, such as boolean models or semantic matching models.
\end{inparaenum}
Future work includes analysis of the effect of various pre-processing steps (e.g., signature removal, thread merging, phrase-level tokenization), a study of the effect of training/evaluating query formulation models on the email of a single organization to measure the effect of organization-specific properties and the incorporation of the social graph in the email domain. In addition, structured queries with operators searching different fields (e.g., recipients, subject) can improve performance. Finally, we assumed a single model for all mailboxes; per-mailbox personalized models are likely to generate better queries.
\smallskip
\begin{spacing}{1}
\noindent\small
\textbf{Acknowledgments.}
The authors would like to thank Maarten de Rijke and the anonymous reviewers for their valuable comments.
The first author was supported, before and after his employment at Microsoft, by
the Bloomberg Research Grant program
and
the Google Faculty Research Award scheme.
All content represents the authors' opinions, which is not necessarily shared or endorsed by their respective employers and/or sponsors.
\end{spacing}

\bibliographystyle{abbrvnat}
\bibliography{replywith} 

\end{document}